\documentclass[11pt]{article}
\usepackage{euscript}
\usepackage{amssymb}
\usepackage{amsfonts}
\usepackage{amsbsy}
\usepackage{amsmath}
\usepackage{epsfig}
\usepackage{amsthm}
\usepackage{amscd}
\usepackage{amstext}

\usepackage{mathrsfs}
\usepackage{color}

\usepackage[pdftex]{hyperref}

\newcommand{\be}{\begin{equation}}
\newcommand{\ee}{\end{equation}}
\newcommand{\bea}{\begin{eqnarray}}
\newcommand{\eea}{\end{eqnarray}}

\def\IC{\mathbb{C}}

\def\IZ{{\mathbb{Z}}}
\def\IR{{\mathbb{R}}}

\def\ICP{\mathbb{CP}}
\def\IRP{\mathbb{RP}}

\def\CN{{\cal N}}
\def\CM{{\cal M}}
\def\CZ{{\cal Z}}
\def\CO{{\cal O}}

\def\sm#1{{\mbox{$#1$}}}

\begin{document}

%\begin{flushright}
%arXiv:yymm.nnnn [hep-th]
%\end{flushright}

\vskip2cm

\begin{center}

\vspace{1cm} { \huge {\bf Orientiholes}}

\vspace{1cm}

Frederik Denef$^{1,2}$, Mboyo Esole$^1$ and Megha Padi$^1$

\vspace{0.8cm}

{\it $^1$ Jefferson Physical Laboratory, Harvard University,
\\
17 Oxford Street, Cambridge, MA 02138, USA \\
\vskip4mm
$^2$ Instituut voor Theoretische Fysica, KU Leuven, \\
Celestijnenlaan 200D, B-3001 Leuven, Belgium \\ }

\vskip1cm

\end{center}

\begin{abstract}

By T-dualizing space-filling D-branes in 4d IIB orientifold compactifications along the three non-internal spatial directions, we obtain black hole bound states living in a universe with a gauged spatial reflection symmetry. We call these objects orientiholes. The gravitational entropy of various IIA orientihole configurations provides an ``experimental'' estimate of the number of vacua in various sectors of the IIB landscape. Furthermore, basic physical properties of orientiholes map to (sometimes subtle) microscopic features, thus providing a useful alternative viewpoint on a number of issues arising in D-brane model building. More generally, we give orientihole generalizations of recently derived wall crossing formulae, and conjecture a relation to the topological string analogous to the OSV conjecture, but with a linear rather than a quadratic identification of partition functions.

\end{abstract}

\newpage

\setcounter{tocdepth}{2}
\tableofcontents
\newpage

%\onehalfspacing

\section{Introduction}
D7-branes and their bound states with lower dimensional D-branes play a central role in modern string phenomenology, including moduli stabilization \cite{Kachru:2003aw,Douglas:2006es}, GUT model building \cite{Beasley:2008dc,Beasley:2008kw,Heckman:2008rb,Blumenhagen:2008zz,Blumenhagen:2006ci}, and models of inflation such as \cite{Baumann:2007ah,Chen:2009nk}. The vacuum degeneracy arising from the D-brane sector moreover vastly dominates that arising from the bulk \cite{Collinucci:2008pf,Denef:2008wq}. Getting a systematic understanding of this sector in a compact setting is an important but difficult task. Explicit enumeration of vacua is computationally intractable. In \cite{Douglas:2003um}, inspired by \cite{Bousso:2000xa}, a program was proposed to analyze large classes of string vacua simultaneously by statistical methods. Basic formulae were developed in \cite{fluxcounting,opencount} to estimate numbers of vacua in specified regions of parameter space, without explicit enumeration. These formulae are approximate, becoming accurate in certain large D-brane charge limits. Tadpole cancelation constraints fix these charges to specific values in actual compactifications, typically in a ``mesoscopic'' range, where the counting formulae are never deep in their asymptotic region of validity. In fact, in the specific problem of counting D7 worldvolume vacua, one generically ends up in a regime where the formulae developed so far are at best poor approximations \cite{Denef:2008wq}. One is therefore led to ask if this situation can be improved.

One idea is the following. In the weak string coupling limit, i.e.\ classically, the internal configuration spaces of space-filling and spatially localized wrapped D-brane systems are identical. Their interpretation is very different though: The former represent vacua, the latter give rise to microstates of particle-like objects, which become black holes at sufficiently large values of the string coupling. Because of this, the problem of counting supersymmetric vacua and counting supersymmetric black hole microstates are in essence mathematically identical, if we take ``counting'' to mean computing the appropriate Witten index, which can be computed at arbitrarily weak coupling. It is essentially the Euler characteristic of the D-brane configuration moduli space. This is in many ways a sensible definition of the notion of counting vacua, as advocated in \cite{Douglas:2003um} and further discussed below in section \ref{sec:countingD7}.

Thus, in principle, we could effectively count vacua in the lab, by producing a mesoscopic BPS black hole with the appropriate charges, and subsequently measuring its entropy! Of course we don't quite live in the right universe to actually do this experiment, but we can perform it at a theoretical level: we can construct the appropriate black hole solutions and compute their entropy by applying the Bekenstein-Hawking entropy formula or a refinement thereof. This is in general a relatively simple task, much simpler in any case than microscopically deriving the numbers. Rather than derive estimates ourselves, we let gravity do the work for us!

In actuality, the mapping is a little more complicated, and also a little more interesting. Physically, what maps space-filling and point-localized D-branes to each other is T-duality. Start with a type IIB Calabi-Yau orientifold compactification of O3/O7 type, containing space-filling D7- and D3-branes, at vanishingly small string coupling. Now imagine we compactify space itself (the ``visible'' space in which we live) on a 3-torus, and T-dualize along the three spatial directions, to a type IIA theory. This maps space filling D7- and D3-branes to space localized D4- and D0-branes. In addition, the orientifold planes become localized too, the original orientifold involution getting extended by the spatial inversion $\vec x \to - \vec x$, which has eight fixed points on the 3-torus.  The total charge of the resulting configuration will be zero, as required by tadpole cancelation on the IIB side and by the compactness of space. We may decompactify space again to $\IR^3$, with an orientifold plane at the origin. In this noncompact setting, the total charge need not be zero, and we can consider arbitrary sets of RR charged particles obtained by wrapping D-branes on internal cycles, as long as they respect the orientifold symmetry, including spatial inversion.

When we increase the gravitational coupling sufficiently, these configurations will turn into black hole systems. The most general stationary BPS solutions will be multicentered black hole bound states similar to those in the parent (unorientifolded) $\CN=2$ theory \cite{Denef:2000nb,Behrndt:1997ny,LopesCardoso:2000qm},
but now subject to the orientifold projection constraints. We call these orientifolded black holes \emph{orientiholes}. As we will see, various physical properties of orientiholes find direct microscopical interpretations. Here are a few examples. A particular $\IZ_2$ torsion charge which can be measured concretely by an Aharanov-Bohm type experiment corresponds to a delicate uncanceled ``K-theory tadpole'' on the IIB side, which manifests itself as a subtle gauge anomaly in certain probe gauge theories.
The intrinsic field angular momentum generated by a pair of orientiholes corresponds to an index counting open strings stretched between the corresponding branes, properly taking into account the orientifold projection. Stability of orientihole bound states maps to D-term stability of orientifold vacua. For systems whose constituents are simple particles rather than black holes, an exact map between Witten indices of multiparticle abd D-brane moduli space quantum mechanics can be established, generalizing results of \cite{Denef:2002ru,Denef:2007vg}.

Although we arrived at them because of their relation to IIB orientifold vacua, orientiholes are quite interesting in their own right, independent of any applications to vacuum statistics. In particular they lead to nontrivial generalizations of recently proposed wall crossing formulae for BPS indices. Most notably perhaps, we conjecture a relation to the topological string similar to the ``$\CZ_{\rm BH} = |\CZ_{\rm top}|^2$'' relation conjectured by Ooguri, Strominger and Vafa in \cite{Ooguri:2004zv}, except that now the relation is \emph{linear}; schematically: $\CZ_{\rm OH} = \CZ_{\rm top}$. Morally speaking this is because orientifolding cuts out real slices and identifies complex conjugates.

In section 2 we derive the correct O4/O0 and O6/O2 IIA orientihole projection conditions by spatially T-dualizing the standard IIB O7/O3 resp.\ O9/O5 orientifold transformations. In section 3 we derive the general form of orientihole solutions and investigate their properties. In section 4 we define an orientihole index, derive a wall crossing formula and formulate a conjectural relation to the topological string. In section 5 we give some applications to counting D7 vacua. Finally in section 6 we discuss directions for future research.

\section{From IIB Vacua to Orientiholes} \label{sec:vactoOr}

In the following we describe how spatial T-duality maps IIB orientifold compactifications with space-filling D-branes and O-planes to IIA compactifications with spatially \emph{localized} D-branes and O-planes.\footnote{We could similarly start in IIA and go to IIB, but we will not do so in this paper.} Since string compactifications with O-planes localized in points of the visible universe are not commonly considered, we detail the constraints on the massless bosonic fields such a setup entails. In this section we will work in the weak coupling limit $g_s \to 0$. In the next section we will describe the corresponding supergravity solutions which arise when $g_s$ is increased -- these are the orientiholes.

\subsection{T-duality}

Consider type IIB string theory compactified on a Calabi-Yau three-fold $X$ characterized by  a unique holomorphic three-form $\Omega^{3,0}$ and a K\"ahler form $J$.
We are interested in a $\mathbb{Z}_2$ orientifold and therefore we assume that the Calabi-Yau $X$ admits a  holomorphic involutive isometry  $\sigma:X\rightarrow X$  preserving the holomorphic three-form up to a sign \cite{Acharya:2002ag,Brunner:2003zm}:
\begin{equation}
\sigma^2=1, \quad\sigma^* J=J, \quad  \sigma^* \Omega^{3,0}=(-)^{\epsilon+1} \Omega^{3,0}
\quad \hbox{where }\quad \epsilon=0,1.
 \end{equation}
From the action of $\sigma$ on $\Omega^{3,0}$ we see that when $\epsilon=0$, an odd number of holomorphic coordinates gets inverted, while when $\epsilon=0$, an even number gets inverted, thus giving rise to the following types of O-planes:
\begin{equation}
 \epsilon = 0 \quad \Leftrightarrow \quad {\rm O3/O7} \, , \qquad \epsilon = 1 \quad \Leftrightarrow \quad {\rm O5/O9}.
\end{equation}
We denote by $\Omega$ the  worldsheet parity operator  which takes left-moving oscillators to right-moving oscillators and vice versa.\footnote{Besides denoting worldsheet orientation inversion and the holomorphic 3-form, further on in this paper, $\Omega$ will also be used to represent the symplectic section of special geometry, and the index of BPS states. We really like this letter a lot.} The  space-time fermion number   in the left-moving sector  will be denoted by $F_L$. The total orientifold action $\tau$ is then \cite{Acharya:2002ag,Brunner:2003zm}:
\begin{equation}
  \tau = \Omega \, (-1)^{(\epsilon+1) F_L} \, \sigma^* \, ,\qquad \epsilon=0,1.
\end{equation}
The inclusion of $(-1)^{F_L}$ when $\epsilon=0$ ensures that the orientifold symmetry is involutive on space-time fermions, i.e.\ $\tau^2=1$.
%It   is determined by the action of the involution $\sigma$ on the holomorphic three-form  ($\sigma^*\Omega^{3,0}=(-)^{\epsilon+1}\Omega^{3,0}$)  and distinguishes between $O3/O7$ and $O5/O9$ orientifolds.
%Explicitly, we have
%\begin{eqnarray}
%O3/O7: &&  \epsilon=0, \quad \sigma^*J=J,\quad\sigma^*\Omega^{3,0} =-\Omega^{3,0},\quad \tau=\Omega\sigma (-1)^{F_L} ,\\
%O5/O9: && \epsilon=1, \quad \sigma^*J=J, \quad\sigma^*\Omega^{3,0} = +\Omega^{3,0},\quad \tau = \Omega \sigma.
%\end{eqnarray}

Our ten dimensional space-time is $M_{10} = \IR^{1,3} \times X$. Let $\vec{x}=(x^1, x^2, x^3)$ the spatial coordinates of $\IR^{1,3}$. We first compactify the three spatial coordinates $\vec x$ on a 3-torus. Next we T-dualize along all three directions, and decompactify the resulting dual 3-torus again, thus ending up in type IIA on $\IR^{1,3} \times X$. The T-duality action $T_{1,2,3}$ along the three directions of $\vec x$ transforms the orientifold transformations as \cite{Polchinski:1996fm,Dabholkar:1996pc}:
\begin{equation}
\sigma\overset{T_{1,2,3}}{\longrightarrow}  \mathscr{P}\sigma, \quad \Omega\overset{T_{1,2,3}}{\longrightarrow}   (-1)^{F_L}  \Omega  ,
\end{equation}
where $\mathscr{P}$ is a reflection in all the three spatial directions $x^1,x^2,x^3$ (now of the T-dual space):
\begin{equation}
\mathscr{P}: \vec x\to -\vec x.
\end{equation}
A type IIB space-time filling  O$p$-plane wrapping a holomorphic submanifold $S$ of the Calabi-Yau $X$ becomes in type IIA a transverse  O$(p-3)$-plane  located at the fixed loci of the involution  $\mathscr{P} \sigma$. That is, it wraps the  holomorphic-submanifold $S$  of the Calabi-Yau three-fold  $X$ and is located at the origin $\vec x=0$ of the universe. The total orientifold action is now given by
\begin{equation} \label{tauprime}
 \tau' = \Omega \, (-1)^{\epsilon F_L} \, \sigma^* \mathscr{P}^* \, .
\end{equation}
where for $\epsilon=0$ we get an $O0/O4$ at the origin and for $\epsilon=1$ an $O2/O6$.

\subsection{Constraints on IIA Bosonic Massless Spectrum} \label{sec:constrIIA}

The massless fields in the effective $d=4$, $\mathcal{N}=2$ supergravity  will be constrained by the orientifold projection. The action of the operators $\Omega$ and $(-1)^{F_L}$  on the ten-dimensional fields of Type II supergravity  \cite{Brunner:2003zm,Grimm:2004uq} is reviewed  in
  Table~\ref{Table.OmegaFL}.
   \begin{table}[!htb]
\begin{center}
\begin{tabular}{|c|c|c|c|c|c|c|c|c|}
\cline{2-9}
\multicolumn{1}{c|}{}& $\phi$ & $g$ & $B$ & $C^{(0)}$ & $C^{(1)}$ & $C^{(2)}$ & $C^{(3)}$ & $C^{(4)}$ \\
\hline
$(-1)^{F_L}$ & $+$  & $+$ & $+$ & $-$ & $-$ & $-$ & $-$ & $-$\\
\hline
$\Omega$ & $+$  & $+$ & $-$ & $-$ & $+$ & $+$ & $-$ & $-$\\
\hline
\end{tabular}
\caption{Parity of the different fields under the actions of $(-1)^{F_L}$ and $\Omega$.
The scalar field $\phi$ is the dilaton, $g$ is the metric, $B$ is the NS-NS  two-form  and $C^{(m)}$ is a  RR $m$-form.
\label{Table.OmegaFL}}
\end{center}
\end{table}
The ten-dimensional fields of type IIA supergravity reduce to four-dimensional fields (see e.g.\ \cite{Grimm:2004uq}):
\begin{align}
\phi&=\phi(x), \qquad
J:= - i g_{m\bar n} dz^m \wedge d\bar z^{\bar n} = v^A(x) \, D_A , \qquad \delta g_{mn}=\delta \bar{z}^k(x) \, \bar\xi_{k,\bar p\bar q(m} \Omega^{\bar p\bar q}{}_{n)} \, ,  \\
B&=b^A (x) \, D_A \, + \, B(x) , \quad
C^{(1)} ={\cal A}^0(x), \quad
C^{(3)} ={\cal A}^A(x) \, D_A+\zeta^K(x) \, \alpha_K-\tilde\zeta_K(x) \, \beta^K.
\end{align}
Here, $m,n,p,q$ are holomorphic coordinate indices on $X$, $\Omega$ is the holomorphic 3-form, $J$ is the K\"ahler form, $\{D_{A}\}_{A=1,\ldots,h^{1,1}}$ is a basis of (harmonic representatives of) $H^{1,1}(X)$, $\{ \xi_{k} \}_{k = 1...h^{2,1}}$ a basis of $H^{2,1}(X)$, and $\{\alpha_K,\beta^K\}_{K=1,\ldots,h^{2,1}+1}$ a symplectic basis of $H^3(X)$.
The complex variables $z^k$ are the complex structure moduli.
%In the expression of $\delta g_{m n }$,
%$\Omega^{\bar m \bar l}{}_n(y)$ is obtained from  holomorphic three-form by raising its two first indices with the inverse of the K\"ahler-metric.
The moduli $b^A$ and $v^A$ parametrize the complexified K\"ahler form
\begin{equation} \label{tdef}
t =B+iJ=t^A D_A=(b^A+i v^A)D_A.
\end{equation}
Together $t^A$ and $z^k$ form the geometric moduli of the Calabi-Yau manifold.

Only field configurations invariant under the total orientifold transformation $\tau'$ defined in (\ref{tauprime}) survive. The action of $\Omega$ and $(-1)^{F_L}$ on the 4d fields descends simply from the action on the 10d fields.  Since $\sigma$ is an involution, it can only have eigenvalues $\pm 1$. Denote the $\pm1$ eigenspace of $H^{1,1}$ under $\sigma^*$ by $H^{1,1}_{\pm}$. Finally, $\mathscr{P}^*$ gives a minus sign for each spatial index. Putting this all together, if the net intrinsic effect of $\tau'$ is a minus sign, then we know that the field must have odd extrinsic parity, that is, be an odd function under $ \vec x\mapsto -\vec x$. We summarize the results for an $O4/O0$ type orientifold in Table \ref{Table.ModuliAIIOmegaSigmaR}, and for an $O6/O2$ type in Table~\ref{Table.ModuliAIIOmegaSigmaRO6}. Notice that unlike in space-filling orientifold theories, \emph{all} of the parent $\CN=2$ fields still appear in the 4d theory. In particular, KK reduction and spatial T-duality do not commute.

As a concrete example, let us focus on the gauge fields ${\cal A}^A_{\mu}$ ($\mu=0,1,2,3)$ in the $\epsilon=0$ case. These fields descend from $C^{(3)}$ so they are odd under $\Omega$. The time component ${\cal A}^A_0$ is intrinsically even under $\mathscr{P}^*$ and the spatial components ${\cal A}^A_i$  ($i=1,2,3$) are intrinsically odd. They also transform under $\sigma^*$ depending on which subspace of $H^{1,1}$ they correspond to. Let us label them ${\cal A}^{A+}_{\mu}$ and ${\cal A}^{A-}_{\mu}$  accordingly.  Then,  we find that ${\cal A}_0^{A+} (t,\vec{x}) = - {\cal A}_0^{A+} (t,-\vec{x})$, i.e. the ${\cal A}_0^{A+}$ field must have odd extrinsic parity. Likewise, ${\cal A}_i^{A+}$ must have even extrinsic parity. Note that for gauge fields whose time component is odd, the total associated electric charge must vanish. Similarly, for gauge fields whose spatial components are odd, the total magnetic charge must vanish. Furthermore, charges located at $\vec{x}$ and $-\vec{x}$ must exactly cancel or equal each other, depending on whether the total charge of the kind under consideration vanishes or not.

\begin{table}[!htb]
\begin{center}
\begin{tabular}{||c||cc||cc||cc||}
\hline
\hline
Multiplets & \multicolumn{2}{c||}{IIA}& \multicolumn{4}{|c||}{Condition to survive  $\tau'=\Omega\sigma^* \mathscr{P}^*$}\\
\cline{4-7}
\cline{4-7}
 & &&  \multicolumn{2}{|c||}{Extrinsic $\mathscr{P}$-Even} &\multicolumn{2} {c||}{Extrinsic $\mathscr{P}$-Odd} \\
\hline
gravity  & $1$ &$(g_{\mu\nu},{\cal A}^0_\mu)$ & $1$& $(g_{00}, g_{ij}, {\cal A}^0_0)$ & $1$ & $(g_{0i},{\cal A}^0_i)$ \\
\hline
vector  &   $h^{1,1}$& (${\cal A}^A_\mu,t^A)$  &    $h^{1,1}_-$ &  $({\cal A}^{A-}_0, b^{A-})$ & $h^{1,1}_- $ & $({\cal A}^{A-}_i,  v^{A-})$  \\
  & &  & $ h^{1,1}_+$ &   $({\cal A}^{A+}_i, v^{A+})$ &  $h^{1,1}_+$ & $({\cal A}^{A+}_0,b^{A+})$\\
\hline
hyper & $h^{2,1}$ & $(z^k,\zeta^k,\zeta_k)$ & $h^{2,1}_-$ & $(z^{k-},\zeta^{k-},\zeta_{k-})$ & $ h^{2,1}_+$ & $(z^{k+}, \zeta^{k+},\zeta_{k+}) $\\
\hline
(double) tensor & $1$ & $( B_{\mu\nu},\phi,\zeta^0,\zeta_0)$& $1$&  $(B_{0i},\phi)$ & $1$&  $(B_{ij},\zeta^0,\zeta_0)$\\
\hline
\hline
\end{tabular}
\end{center}
\caption{Classification of bosonic matter content of the IIA $\epsilon=0$ (O4/O0) theory. A field $F$ is  ``Extrinsic $\mathscr{P}$-even" if $F(t,\vec{x}) = F(t,-\vec{x})$ and ``Extrinsic $\mathscr{P}$-odd" if $F(t,\vec{x}) = -F(t,-\vec{x})$. }
 \label{Table.ModuliAIIOmegaSigmaR}
\end{table}

\begin{table}[!htb]
\begin{center}
\begin{tabular}{||c||  ll  ||  ll ||  ll  ||}
\hline
\hline
Multiplets
& \multicolumn{2}{c||}{IIA}& \multicolumn{4}{|c||}{Condition to survive
$\tau'=\Omega(-1)^{F_L}\sigma^* \mathscr{P}^*$}\\
\cline{4-7}
\cline{4-7}
 & &&  \multicolumn{2}{|c||}{Extrinsic $\mathscr{P}$-Even}
&\multicolumn{2} {c||}{Extrinsic $\mathscr{P}$-Odd} \\
\hline
gravity & $1$ &$(g_{\mu\nu},{\cal A}^0_\mu)$ & $1$& $(g_{00},
g_{ij}, {\cal A}^0_i)$ & $1$ & $(g_{0i},{\cal A}^0_0)$ \\
\hline
vector  &   $h^{1,1}$& (${\cal A}^A_\mu,t^A)$  &
$h^{1,1}_-$ &  $({\cal A}^{A-}_i, b^{A-})$ & $h^{1,1}_- $ & $({\cal
A}^{A-}_0,  v^{A-})$  \\
  & &  & $ h^{1,1}_+$ &   $({\cal A}^{A+}_0, v^{A+})$ &  $h^{1,1}_+$ &
$({\cal A}^{A+}_i,b^{A+})$\\
\hline
hyper & $h^{2,1}$ & $(z^k,\zeta^k,\zeta_k)$ & $h^{2,1}_+$ & $(z^{k+},\zeta^{k+},\zeta_{k+})$ & $ h^{2,1}_-$ & $(z^{k-}, \zeta^{k-},\zeta_{k-}) $\\
\hline
(double) tensor  & $1$ & $( B_{\mu\nu},\phi,\zeta^0,\zeta_0)$&
$1$&  $(B_{0i},\phi,\zeta^0,\zeta_0)$ & $1$&  $B_{ij}$\\
\hline
\hline
\end{tabular}
\end{center}
\caption{Classification of bosonic matter content of Type IIA $\epsilon=1$ (O6/O2) theory.}
 \label{Table.ModuliAIIOmegaSigmaRO6}
\end{table}

\section{Supergravity Description of Orientiholes}

In the previous section we saw how IIB space-filling D-branes and orientifold planes get dualized into pointlike D-branes and orientifold planes. In the weak string coupling limit, from a four dimensional point of view, these objects are RR-charged points in a flat background. Tadpole cancelation on the IIB side is equivalent to the total charge of the system being zero. For space-filling branes zero total charge is necessary because otherwise the flux lines have nowhere to go. For localized charges in a noncompact space this is no longer an issue, and we are free to consider configurations with arbitrary charges. For sufficiently large charges and away from the strict weak string coupling limit, these configurations of charges will become configurations of black holes. Apart from the obvious single centered black holes sitting on top of the orientifold point at the origin, multicentered bound states of black holes symmetric around the origin are also possible, an orientifolded version of the familiar multicentered  black hole bound states in ${\cal N}=2$ theories \cite{Denef:2000nb,Behrndt:1997ny,LopesCardoso:2000qm,Bates:2003vx,Denef:2007vg}.

In the following we will begin by reviewing $\mathcal{N}=2$ black hole solutions, and then explain how to incorporate the orientifold projections. We discuss the gravitational entropy of these solutions and fix ideas by looking in more detail at a simple example. In particular we will explicitly consider the effect of the negative tension of O-planes. We then go on to discuss angular momentum and decay at marginal stability, as well as attractor flow trees and how they get modified in this context. In the last part we carefully analyze how Dirac charge quantization is affected by orientifolding, and we identify $\IZ_2$ torsional charges that come into existence. This charge can concretely be measured by an Aharonov-Bohm type experiment. Its nonvanishing is T-dual to the presence of a very subtle ``K-theory tadpole'', which usually in model building applications is only detected in an indirect way by finding gauge anomalies in probe brane theories.

\subsection{Review of stationary multicentered $\mathcal{N}=2$ black holes} \label{sec:revN2}

We consider a four dimensional ${\cal N}=2$ supergravity theory consisting of Einstein gravity coupled to $(n_V+1)$  massless Abelian gauge fields ${\cal A}^\Lambda$, $\Lambda=0,1,\cdots, n_V$ and  $n_V$  complex scalar fields    $t^A$ ($A=1,\cdots, n_V$). The hypermultiplets decouple and can be consistently put to any constant value. We work in units with the 4d Newton constant $G_N \equiv 1$.

The lattice $L$ of magnetic-electric charges $\Gamma$ carries a fundamental symplectic product, which in a symplectic basis has the canonical form
\begin{equation} \label{symplprod}
\langle \Gamma,\Delta  \rangle:=\Gamma^\Lambda \Delta_\Lambda-\Gamma_\Lambda \Delta^{ \Lambda}.
\end{equation}
Upper indices denote magnetic and lower indices electric components.

A \emph{single} centered static BPS solution to the equations of motion has a metric of the form
$ds^2 = -e^{2U} dt^2 + e^{-2U} d\vec x^2$, with $U$ and $t^A$ functions of $r=|\vec x|$ only. The BPS equations of motion are \cite{Ferrara:1995ih,Ferrara:1997tw}
\begin{align} \label{singflow1}
\dot{U} &= \mp e^U |Z| ,\\
\dot{t}^A &= \mp 2 e^U g^{A\bar{B}} \partial_{\bar{B}} |Z|, \label{singflow2}
\end{align}
where $g_{A\bar{B}}$ is the metric on the scalar manifold, the dot denotes derivation with respect to $\tau \equiv 1/r$, and $Z(\Gamma,t)$ is the \emph{central charge} of the magnetic-electric charge $\Gamma$ at moduli $t$, expressed in terms of the  (normalized) symplectic section $\Omega(t)$ defining special geometry \cite{Strominger:1990pd} as
\begin{equation} \label{ZZZ}
Z(\Gamma,t)=\langle \Gamma, \Omega(t) \rangle.
\end{equation}
Taking the minus sign in the equations, we get well-behaved black hole solutions with positive ADM mass $|Z|_{r=\infty}$. Taking the positive sign, we get singular, gravitationally repulsive solutions with negative ADM mass $-|Z|_{r=\infty}$. The latter solutions are usually rejected, but we will later interpret them to be the solutions sourced by localized orientifold planes, which are indeed intrinsically singular objects with negative mass. Of course actual orientifold planes will always have microscopically small charges, so corrections can be expected to be important and possibly resolve the singularity. However at large distances from the orientifold point the massless fields will be given by the solution to the above equations.

An alternative way of writing the equation of motion is \cite{Ferrara:1995ih,Denef:2000nb}
\begin{equation} \label{floweq2}
 2 \,\frac{d}{d\tau} {\rm Im}\bigl[e^{-U-i \alpha} \Omega(t)\bigr]  = -\Gamma \, ,
\end{equation}
which can trivially be integrated. The position dependent phase $\alpha$ determines which $\CN=1$ subalgebra of the parent $\CN=2$ is locally preserved. It is fixed by the boundary condition
\begin{equation} \label{alphadef}
 e^{i \alpha} = \pm \frac{Z}{|Z|}
\end{equation}
at spatial infinity, which then in fact holds everywhere, as can be seen by taking the symplectic product of $\Gamma$ with (\ref{floweq2}). The choice of sign here corresponds to the choice of sign in (\ref{singflow1}).

General, multicentered, stationary BPS solutions have a metric of the form
\begin{equation}
 ds^2=-e^{2U}(dt+\omega_i dx^i)^2+ e^{-2U}d\vec x^2,
\end{equation}
where $U$ and $\omega$ depend on $\vec x$. Denote by $\vec x_s$ and  $\Gamma_s=(\Gamma_s^\Lambda,\Gamma_{s,\Lambda})=(P^\Lambda_s, Q_{s,\Lambda})$  the position  and   the  (magnetic,electric) charges  of the $s$-th black hole with respect to the abelian gauge fields  ${\cal A}^\Lambda$. The generalization of (the integrated version of) (\ref{floweq2}) is
\begin{equation} \label{OmH}
 2 e^{-U} {\rm Im}\bigl[e^{-i \alpha} \Omega(t)\bigr] = - H \, , \qquad H:=( H^\Lambda, H_\Lambda)=\sum_s \Gamma_s\tau_s +h \, ,  \qquad \tau_s:=\frac{1}{|\vec x-\vec x_s|}.
\end{equation}
The constant term in the harmonic function $H$ is
\begin{equation}
\label{hDefinition}
h=-2\,\mathrm{Im}\left(e^{-i \alpha} \Omega(t) \right)|_{r=\infty},
\end{equation}
where $\alpha|_{r=\infty}$ is again given by (\ref{alphadef}), with $Z$ now the total central charge $Z(\Gamma)=Z(\sum_s \Gamma_s)$ evaluated at $r=\infty$. Again $\alpha$ determines the locally preserved supersymmetry.

The gauge field ${\cal A}$ is given by
\begin{equation}
 {\cal A}^\Lambda = 2 e^U {\rm Re} \bigl[ e^{-i \alpha} \Omega^\Lambda \bigr] (dt + \omega) + {\cal A}^\Lambda_{\mathscr{D}}
\end{equation}
The one-forms $\omega$ and  ${\cal A}^\Lambda_{\mathscr{D}}$ are respectively solutions to the  equations
\begin{equation}
d\omega=\star \langle dH, H\rangle, \qquad d{\cal A}^\Lambda_{\mathscr{D}}=-\star dH^\Lambda \, ,
\end{equation}
where $*$ is the flat $\IR^3$ Hodge star. The one-form ${\cal A}^\Lambda_{\mathscr{D}}$ is  the vector potential  for a system of Dirac magnetic monopoles  of charge $p^\Lambda_s$ located at $\vec x_s$.

Solving equation (\ref{OmH}) for the metric warp factor $e^{2U}$ and the moduli $t^A$ is still nontrivial. However, once the single centered BPS Bekenstein-Hawking entropy $S(P,Q)$ is known as a function of the charge $(P,Q)$, everything else follows \cite{Bates:2003vx}. The BPS entropy function itself is defined in general as
\begin{equation}
 S(\Gamma) = \frac{{\rm Area}}{4} = \pi \min_t |Z(\Gamma,t)|^2 \, .
\end{equation}
More precisely, the minimum is the minimum we reach by following the gradient flow (\ref{singflow1}). In some cases this can depend on the ``area code'' (basin of attraction) of the background moduli $t|_{r=\infty}$ \cite{Moore:1998pn,Denef:2001xn}. If the value zero is reached at a regular interior point of the moduli space, no single centered solution exists \cite{Moore:1998pn}.

The solutions for the metric, gauge fields and scalar fields are then given by
\begin{equation} \label{Sigmadef}
e^{-2U} = \Sigma := \frac{1}{\pi} S(H^\Lambda,H_\Lambda) \, ,
\qquad
{\cal A}^\Lambda=\frac{\partial \log   \Sigma}{\partial H_\Lambda}  \  (dt+\omega)+{\cal A}^\Lambda_{\mathscr{D}}, \qquad
t^A=\frac{H^A-i \frac{\partial \Sigma}{\partial H_A}}{H^0-i \frac{\partial \Sigma}{\partial H_0}}.
\end{equation}
In particular $\Sigma(h)=1$ for $h$ as defined in (\ref{hDefinition}).  The equation $d\omega=\star \langle dH, H\rangle$ implies an important  integrability condition
 $d^2 \omega=0$ which can be written as
$\langle \triangle H, H\rangle=0$.
  Since $\triangle \tfrac{1}{|\vec x-\vec x_s| }=-4\pi \delta^3(\vec x-\vec x_s)$,  the integrability condition  becomes a constraint on the relative  distances between the positions of the centers (in $\mathbb{R}^3$)  except  when the charges are mutually local  (i.e., for all $r,s$: $\langle \Gamma_s, \Gamma_r \rangle=0$). More precisely  for each     $s$ we have
   \begin{equation}\label{IntegrabilityConditions}
 \sum_r\frac{ \langle \Gamma_s, \Gamma_r\rangle}{|\vec x_s -\vec x_r|}
+  \langle \Gamma_s, h\rangle=0.
  \end{equation}

The integrability conditions are a necessary but not sufficient condition on the existence of solutions. One also needs to check if the entropy function $\Sigma$ in (\ref{Sigmadef}) is everywhere nonvanishing. This is in general difficult to do. In \cite{Denef:2000ar,Denef:2007vg} the conjecture was formulated that the existence of solutions is equivalent to the existence of so called \emph{attractor flow trees}. An attractor flow tree is built out of single attractor flows. The tree starts at the background value of the moduli and terminates at the attractor points of the constituent charges. Each
edge $E$ of an attractor flow tree is given by a single charge attractor flow for some charge $\Gamma_E$. Charge and energy is conserved at the vertices, i.e.\ for each vertex splitting $E \to (E_1,E_2)$,
$\Gamma_E = \Gamma_{E_1} + \Gamma_{E_2}$ and
$|Z(\Gamma_E)|=|Z(\Gamma_{E_1})| + |Z(\Gamma_{E_2})|$. The last
condition is equivalent to requiring the vertices to lie on a line
of marginal stability: $\arg Z(\Gamma_{E_1}) = \arg
Z(\Gamma_{E_2})$. Flow trees can be thought of as giving a recipe for adiabatically assembling or disassembling BPS bound states. We refer to section 3.2.2 of \cite{Denef:2007vg} for a more detailed discussion.

Finally, the solutions have intrinsic angular momentum, given by
\begin{equation}\label{AngularMomentum}
 \vec J=\frac{1}{2}\sum_{s<r}\langle \Gamma_s, \Gamma_r \rangle \frac{\vec x_s-\vec x_r}{|\vec x_s-\vec x_r|}.
\end{equation}
This is the familiar classical field angular momentum sourced by a system of dyons. For a system of two mutually nonlocal BPS particles without internal degrees of freedom beyond the universal ``center of mass'' fermionic oscillator states filling out a hypermultiplet, the angular momentum quantum number of the \emph{quantum} ground state is
\begin{equation} \label{J2}
 j = \frac{1}{2}|\langle \Gamma_1,\Gamma_2 \rangle| - \frac{1}{2} \, .
\end{equation}
The $-1/2$ correction is due to the spin-magnetic coupling of the particles \cite{Denef:2002ru}.

\subsection{IIA Orientiholes}

\begin{figure}[h]
\begin{center}
\includegraphics[height=6cm]{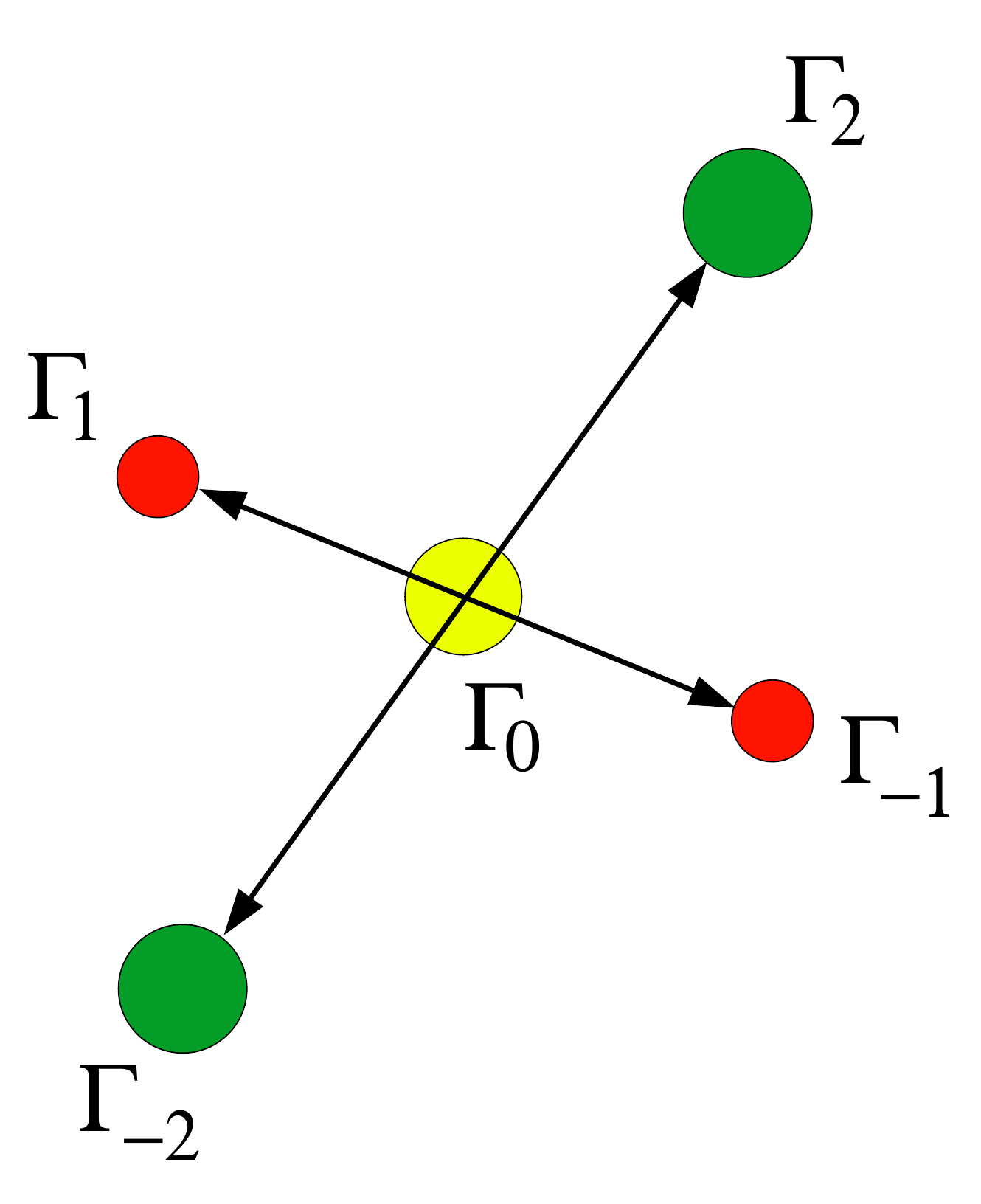}
\end{center}
\caption{An orientihole configuration. \label{orientiholes}}
\end{figure}

We now consider the IIA compactification described in the previous section. Its four dimensional effective field theory is an ${\cal N}=2$ supergravity theory subject to the orientifold projection constraints detailed there. Within this theory we consider a system consisting of an orientifold-invariant charge $\Gamma_0$ at the origin, which includes the orientifold plane charge, plus a set of charges and their images under the orientifold action $\tau'$ defined in (\ref{tauprime}), as in figure \ref{orientiholes}.

For a charge $\Gamma_{+s}$ located at $\vec x_s$, its image charge is denoted ${\Gamma}_{-s}$ and is located at  $-\vec{x}_{s}$. The index   $s=0,\cdots, n$ is such that  $s=0$ corresponds to $\Gamma_0$ located at $\vec x_0=\vec 0$.  It will also be useful to introduce  indices
$a,b=-n,\cdots, 0, \cdots, n$.
The corresponding  stationary  multi-centered solution is given by the harmonic function:
\begin{equation} \label{Horeq}
H=\sum_{a=-n}^n  \Gamma_{a}  \tau_{a} +h,
\end{equation}
where $\tau_{\pm s} = |\vec{x}\pm \vec{x}_s|^{-1}$ and $h=-2\mathrm{Im}\left(e^{-i \alpha} \Omega(t) \right)|_{r=\infty}$ as in (\ref{hDefinition}). In the IIA large volume regime, charges $\Gamma=(P^0,P^{A+},P^{A-},Q_{A+},Q_{A-},Q_0)$ can be thought of as even cohomology classes and correspond to (D6,D4$^+$,D4$^-$,D2$^+$,D2$^-$,D0)-branes via the  decomposition\footnote{In order to have the canonical expression for the symplectic product (\ref{symplprod}), we are defining D0-charge with the opposite sign as the conventions of \cite{Denef:2007vg}.}
\begin{eqnarray}
\Gamma &=& P^0 + P^{A+} D_{A+} + P^{A-} D_{A-} + Q_{A+} \tilde{D}^{A+} + Q_{A-} \tilde{D}^{A-} - Q_0 \, dV \, , \\
&=:& (P^0,P^{A+},P^{A-},Q_{A+},Q_{A-},Q_0) \, .
\end{eqnarray}
Here $dV$ is the unit volume element (i.e.\ $\int_X dV \equiv 1$) and $\tilde{D}^{A\pm}$ a basis of $H^4_\pm$ dual to the basis $D_{A\pm}$ of $H^2_\pm$ introduced in section \ref{sec:constrIIA}. The symplectic product (\ref{symplprod}) can be written more intrinsically as
\begin{equation} \label{intrsympl}
 \langle \Gamma_1,\Gamma_2 \rangle = \int_X \Gamma_1 \wedge \Gamma_2^*
\end{equation}
where $\Gamma^*$ is obtained from $\Gamma$ by inverting the sign of the 2- and 6-form components.

The orientifold transformation maps a charge $\Gamma$ to
\begin{equation}
 \Gamma' := (-)^{\epsilon F_L}\Omega\sigma \Gamma \, ,
\end{equation}
where we recall $\epsilon=0$ for an $O0/O4$ projection and $\epsilon=1$ for the $O2/O6$ case. Explicitly:
\begin{equation} \label{mapsms}
\begin{array}{rrrrrrrrrl}
                         & \Gamma &=( & P^0 & P^+ & P^- & Q_{+} & Q_{-} & Q_{0} &) \\
\epsilon=0 \, \Rightarrow& \Gamma' &=( &-P^0 & P^+ &-P^- &-Q_{+} & Q_{-} & Q_{0} &) \\
\epsilon=1 \, \Rightarrow& \Gamma' &=( & P^0 &-P^+ & P^- & Q_{+} &-Q_{-} &-Q_{0} &) \, .
\end{array}
\end{equation}
Here we suppressed the $A$ indices. Thus we have:
\begin{equation}
 \Gamma_{-s} = \Gamma_s' \, .
\end{equation}
In particular for an invariant charge $\Gamma$ such as the total charge $\sum_a \Gamma_a$ or the charge at the origin $\Gamma_0$, the odd components must vanish:
\begin{equation} \label{orinvQ}
\begin{array}{rrrccccccl}
\epsilon=0 \, \Rightarrow& \Gamma &=( & 0 & P^+ & 0 & 0 & Q_{-} & Q_{0} &) \\
\epsilon=1 \, \Rightarrow& \Gamma &=( & P^0 &0 & P^- & Q_{+} &0 &0 &).
\end{array}
\end{equation}
The symplectic product satisfies
\begin{equation} \label{symplprodrule}
 \langle \Gamma_1',\Gamma_2' \rangle = - \langle \Gamma_1,\Gamma_2 \rangle = \langle \Gamma_2,\Gamma_1 \rangle \, .
\end{equation}
To compute the constant term $h$ in (\ref{Horeq}), we need the symplectic section $\Omega(t)$ at spatial infinity. In the large volume approximation (in which we will work throughout the paper) this is
\begin{eqnarray} \label{Omvalinf}
 \Omega(t)|_{r=\infty} &=& -\frac{e^{b+iv}}{\sqrt{\frac{4}{3}v^3}}|_{r=\infty} \nonumber \\
 &=&- \frac{1}{{\sqrt{\frac{4 (v^+)^3}{3}}}} \left(1 , i v^+ , b^- , \frac{1}{2}(b^-)^2-\frac{1}{2}(v^+)^2, i v^+ b^- ,
 \frac{i}{6}(v^+)^3 - \frac{i}{2}(b^-)^2 v^+ \right)|_{r=\infty} \, .
\end{eqnarray}
Here we used (\ref{tdef}) and the fact that the moduli at infinity are constant and hence subject to the extrinsic $\mathscr{P}$-even orientifold projection conditions of tables \ref{Table.ModuliAIIOmegaSigmaR} and \ref{Table.ModuliAIIOmegaSigmaRO6}, so
\begin{equation}
 v^-|_\infty=0, \qquad b^+|_\infty=0.
\end{equation}
The various products appearing are wedge products, defined in components by
\begin{equation}
 (xy)_A:=D_{ABC}x^B y^C, \qquad xyz:=D_{ABC} x^A y^B z^C \, dV \, ,
\end{equation}
with $D_{ABC}:=\int_X D_A D_B D_C$ the geometric triple intersection numbers. Only triple intersections with an even number of orientifold odd forms can be nonzero. Often, $xyz$ will instead denote the number $\int_X xyz = D_{ABC} x^A y^B z^C$. This should be clear from the context.

From this we can compute the asymptotic central charge $Z(\Gamma,t)=\langle \Gamma,\Omega(t) \rangle|_{r=\infty}$ of any given charge $\Gamma=(P^0,P^+,P^-,Q_+,Q_-,Q_0)$:
\begin{eqnarray}
 Z(\Gamma,t)|_{r=\infty} &=& \frac{1}{{\sqrt{\frac{4 (v^+)^3}{3}}}} \biggl(
 P^+ \frac{1}{2}\left((v^+)^2 - (b^-)^2\right) + Q_- b^- + Q_0 \nonumber \\
 && \qquad  +i\biggl[P^0 \left(\frac{1}{2} (b^-)^2 v^+ - \frac{1}{6}(v^+)^3\right) - P^- v^+ b^- + Q_+ v^+ \biggr] \biggr)|_{r=\infty} \, . \label{Zexpr}
\end{eqnarray}
Now, crucially, \emph{the choice of orientifold projection fixes the preserved asymptotic ${\cal N}=1$ subalgebra within the original ${\cal N}=2$, and therefore the phase $\alpha|_{r=\infty}$ in (\ref{hDefinition}).} This phase will be the phase of the central charge of supersymmetry-preserving D-brane charges. Specifically:
\begin{equation} \label{alphaval}
 \epsilon = 0 \quad \Rightarrow \quad \alpha_\infty=0 \, , \qquad \epsilon = 1 \quad \Rightarrow \quad \alpha_\infty=-\frac{\pi}{2} \, ,
\end{equation}
corresponding to the large volume phases of the central charges of D4$^+$ resp.\ D6-branes. Along a single charge attractor flow with invariant charge $\Gamma$, $\alpha$ will remain constant at this value. For generic multicentered solutions $\alpha$ varies over space in a $\IZ_2$ symmetric fashion.

The phase $\alpha_\infty$ is fixed by the choice of orientifold involution and does not necessarily coincide with the phase of the total charge $\Gamma_{\rm tot}$ of the solution. Indeed if a (non-exotic) orientifold plane is present at the origin and nothing else, the phase will be \emph{opposite} to that of the total central charge, i.e.\ we get the minus sign in (\ref{alphadef}). This is in accordance with the fact that the charge of the orientifold plane is opposite to the charge of a D-brane preserving the same supersymmetry. This is, of course, what makes it possible to supersymmetrically cancel RR charges in orientifold compactifications. As mentioned already in section \ref{sec:revN2} and as we will detail below, this will give rise to a negative ADM mass solution. Only when a sufficient amount of D-branes is added, (\ref{alphadef}) will hold with the plus sign. In general, the mass $M$ of a BPS orientihole solution of charge $\Gamma$ and asymptotic moduli $t$ is
\begin{equation} \label{massformula}
 M = Z(\Gamma,t)  \quad (\epsilon=0) \, , \qquad M = i Z(\Gamma,t) \quad (\epsilon=1) \, .
\end{equation}
Note that these expressions are always real for charges of the form (\ref{orinvQ}).

Using the above values for $\alpha_\infty$ and (\ref{Omvalinf}), we find for the constant term $h=(h^0,h^+,h^-,h_+,h_-,h_0)$ in (\ref{Horeq}):
\begin{equation}
\begin{array}{rcrrccccccl}
\epsilon=0 \, \Rightarrow& h &=&\sqrt{\frac{3}{(v^+)^3}} ( & 0 & v^+ & 0 & 0 & v^+b^- & \frac{1}{6}(v^+)^3 - \frac{1}{2}(b^-)^2 v^+  &) \\
\epsilon=1 \, \Rightarrow& h &=&\sqrt{\frac{3}{(v^+)^3}}( &1 & 0   & b^- & \frac{1}{2}(b^-)^2-\frac{1}{2}(v^+)^2 & 0 & 0 &),
\end{array}
\end{equation}
evaluated at $r=\infty$. Note that this is of the form of the orientifold invariant charges (\ref{orinvQ}), as it should be.

The BPS solutions constructed from the harmonic function (\ref{Horeq}) as reviewed in section \ref{sec:revN2} will automatically satisfy the projection constraints of section \ref{sec:constrIIA}. Explicit solutions for $\omega$ and ${\cal A}_{\mathscr{D}}$ can be found in appendix \ref{app:explicit}.

\subsection{Orientropy}

Evidently, the total leading order Bekenstein-Hawking entropy of an orientihole configuration is the total horizon area in the quotiented spacetime, i.e.\ half the entropy of the corresponding black hole configuration in the $\IZ_2$ covering spacetime:
\begin{equation} \label{orientropyval}
 S_{\rm tot} = \frac{1}{2} \sum_{a=-n}^n S_{\rm BH}(\Gamma_a) = \frac{1}{2} S_{\rm BH}(\Gamma_0) + \sum_{s=1}^n S_{\rm BH}(\Gamma_s) \, .
\end{equation}
In particular for $\epsilon=0$, using (\ref{orinvQ}) and the results of \cite{Shmakova:1996nz} (see also \cite{Moore:1998pn}), we see that for orientifold invariant charges $\Gamma$ such as $\Gamma_0$
\begin{equation}
 S_{\rm BH}(\Gamma) = 2 \pi \sqrt{\frac{\widehat{Q}_0 P^3}{6}} \, , \qquad \widehat{Q}_0 := Q_0 + \frac{Q^A Q_A}{2} \, , \quad Q^A:=(D_{ABC}P^C)^{-1} Q_B \, .
\end{equation}
The corresponding attractor point $t=b+iv$ lies at
\begin{equation}
 b^A = -Q^A , \qquad v^A = \sqrt{\frac{6 \widehat{Q}_0}{P^3}} \, P^A \, .
\end{equation}
As a consistency check, notice that for charges of the form (\ref{orinvQ}) with $\epsilon=0$, this gives $b \in H^2_-$ and $v \in H^2_+$.

For $\epsilon=0$ and charge only at the origin, these expressions are sufficient to obtain fully explicit solutions as reviewed in \ref{sec:revN2}. When $\epsilon=1$ or whenever we have more general multicentered configurations involving also D6-charges, we need a more general formula for the entropy. This is in general not known in closed form even in the large radius approximation. An exception is when the charge is of the special form
\begin{equation}
 \Gamma = p^0 + p D + q D^2 - q_0 D^3 \, ,
\end{equation}
where $D^m \in H^{2m}$, mapping the problem effectively to the one modulus case. Then we have explicitly:
\begin{eqnarray*}
S_{\rm BH}(p^0,p,q,q_0) &=& \frac{\pi D^3}{3} \,{\sqrt{{\Delta}} \, , \qquad {\Delta} = 3\,p^2\,q^2 - 8\,p^0\,q^3
+ 6\,p^3\,q_0 -
        18\,p\,p^0\,q\,{q_0} -
        9\,{{p^0}}^2\,{{q_0}}^2} \\
t(p^0,p,q,q_0) &=& \frac{p\,q + 3\,{p^0}\,{q_0} + i
    {\sqrt{{\Delta}}}}{p^2 -
    2\,{p^0}\,q} \, D \, .
\end{eqnarray*}

\subsection{A simple example} \label{sec:simple}

At this stage it may be useful to consider a concrete example of a single centered orientihole.

%\subsubsection{General one modulus single centered D4-D0 orientihole}

Let $\epsilon=0$ and assume {\it (i)} $H_+^2=H^2$ is one dimensional with basis element $D$, {\it (ii)} all charge is at the origin and given by
\begin{equation}
 \Gamma_0 = \Gamma = p D - q_0 D^3 \, ,
\end{equation}
and {\it (iii)} the modulus at spatial infinity is given by
\begin{equation}
 v|_{r=\infty}=y D \, .
\end{equation}
Then the above constructions give us a metric of the form $ds^2=- \Sigma^{-1}(r) dt^2 + \Sigma(r) d\vec{x}^2$,
where
\begin{equation} \label{SigmaEx}
 \Sigma(r) = D^3 \sqrt{\frac{2}{3} \left( \frac{q_0}{r} + \sqrt{\frac{y^3}{12 \, D^3}} \right)
 \left( \frac{p}{r} + \sqrt{\frac{3}{D^3 y}} \right)^3} \, .
\end{equation}
The moduli fields are given by
\begin{equation}
 b(r)=0 \, , \qquad v(r) = v^+(r) = \sqrt{6} \, \sqrt{\frac{\frac{q_0}{r} + \sqrt{\frac{y^3}{12 \, D^3}}}{\frac{p}{r} + \sqrt{\frac{3}{D^3 y}}}} \, D \, .
\end{equation}
When $q_0,p>0$ these expressions are manifestly sensible everywhere, describing a regular orientihole. However if for example the charge is purely due to an O4$^-$ plane, which has $p<0$, we get a singular solution, as announced earlier.

\begin{figure}[h]
\begin{center}
\includegraphics[height=6cm]{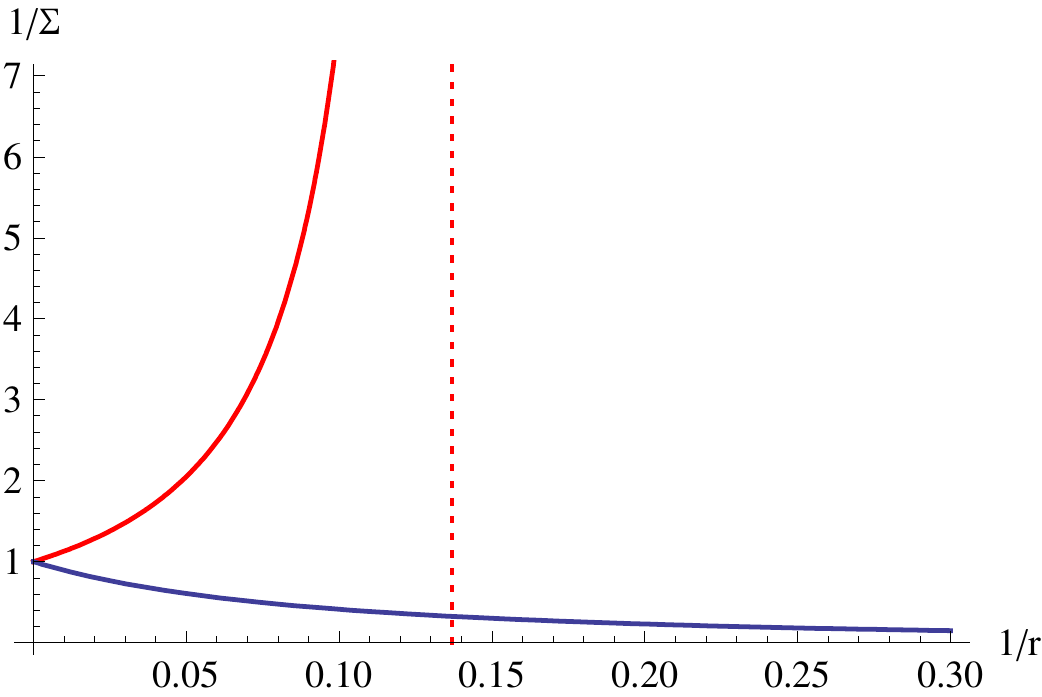}
\end{center}
\caption{Warp factor $1/\Sigma$ as a function of inverse radius $1/r$. The red line diverging at $1/r \approx 0.137$ corresponds to the O4$^-$ example of the text, with $y=5$. For comparison, the blue line converging to zero corresponds to the same charges but with opposite signs for $p$ and $q_0$, giving a regular solution. The upward slope of the O4$^-$ warp factor implies gravitational repulsion. \label{warp}}
\end{figure}

%\subsubsection{O4 solution for $X_{4,1,1,1,1}[8]$}

To see this more concretely, let $X$ be the Calabi-Yau defined as the zero locus of a polynomial of degree $8$ in the weighted projective space $\mathbb{CP}^4_{4,1,1,1,1}$. This was the main example considered in \cite{Collinucci:2008pf} in a IIB context.
We  denote  by $[z_0,z_1,z_2,z_3,z_4]$  the  coordinates  of respective weights $(4,1,1,1,1)$ in $\mathbb{CP}^4_{4,1,1,1,1}$. Then we can take $X$ to be defined by
\begin{equation} \label{CYeq}
 z_0^2 = h(z_1,z_2,z_3,z_4) \, ,
\end{equation}
where $h$ is homogeneous of degree 8. We take the IIA orientifold involution to be
\begin{equation}
 \tau'=\Omega \sigma^* \mathscr{P}^* \, , \qquad \sigma:z_0 \mapsto -z_0 \, , \quad \mathscr{P}:\vec{x} \mapsto -\vec{x} \, .
\end{equation}
The fixed point set of  $\sigma \mathscr{P}$ defines an O4$^-$-plane localized at the origin of space:
\begin{equation}
 O4:z_0=0 \, , \vec{x}=0 \, .
\end{equation}
For this example we have $h^2_+(X) = h^2(X) = 1$ and we can take as basis element $D$ of $H_+^2(X,\IZ)$ the cohomology class Poincar\'e dual to the hyperplane $x_1=0$, which has triple intersection product $D^3=2$. Then the class of the O4 is $4D$, so its D4-charge equals $-4D$. Its D0-charge can be computed\footnote{see \cite{Collinucci:2008pf}, keeping in mind that the charge vector of the O4 has an overall factor of $\frac{1}{8}$ compared to the O7 case analyzed there.} to be $-\frac{\chi(O7)}{48} = -\frac{19}{3}$. Thus its total charge can be written as
\begin{equation} \label{GamO4}
 \Gamma_{O4} = -4D + \frac{19}{6} D^3 \, .
\end{equation}
If this is the only charge present, (\ref{SigmaEx}) becomes
\begin{equation} \label{SigmaExEx}
 \Sigma(r) = 2 \sqrt{\frac{2}{3} \left( -\frac{19}{6 \, r} + \sqrt{\frac{y^3}{24}} \right)
 \left( -\frac{4}{r} + \sqrt{\frac{3}{2y}} \right)^3} \, .
\end{equation}
This is plotted in fig.\ \ref{warp}. Notice that the metric warp factor $\Sigma(r)$ \emph{decreases} with decreasing $r$. This means the object has negative energy, as expected for an orientifold plane. When the asymptotic CY size modulus $y$ is sufficiently large,\footnote{When it is sufficiently small, the other factor hits zero first and the moduli flow to $v=0$. The change in behavior occurs when crossing the local minimum of $|Z|$, the ``repulsor'' point.} $\Sigma$ first hits zero (coming from $r=\infty$) at
\begin{equation}
 r=r_0=4 \sqrt{\frac{2}{3}} \sqrt{y} \, .
\end{equation}
This locus is a curvature singularity with infinite blueshift. Moreover $v \to \infty$ when $r \to r_0$. Thus what we get is an \emph{inverted} attractor flow, the ``wrong sign'' branch of (\ref{singflow1}).

The singularity which manifests itself at the scale $r_0$ is clearly physically unacceptable. We are working in four dimensional Planck units, so we see that by taking $y$ large, the breakdown can be made to happen at arbitrarily large $r_0$ in Planck units. This might seem worrisome at first sight. However recall that the string length $\ell_s$ is related to the Planck length $\ell_4$ by
\begin{equation}
 \ell_s \sim \frac{\sqrt{{\rm vol}(X)}}{g_s} \, \ell_4 \sim \frac{y^{3/2}}{g_s} \, \ell_4 \, ,
\end{equation}
so at fixed $g_s$, the ratio $r_0/\ell_s \sim g_s/y \to 0$ when $y \to \infty$. Hence at fixed (small) $g_s$ stringy effects will always be important at the scale set by $r_0$. If we take at the same time $g_s \to \infty$ we have to switch to M-theory, and then we find $r_0 \sim R/y$ with $R$ the radius of the M-theory circle. So in this case the KK length scale is much larger than the scale set by $r_0$. In either case, the singularity can be expected to get resolved by physics ignored in the 4d supergravity description.\footnote{How such a resolution works in detail is known for the O6$^-$ in $\IR^{1,9}$: Naively uplifting the singular IIA solution to M-theory by neglecting KK modes gives a singular, negative mass Taub-NUT. This is valid at large distances from the origin. The exact solution eleven dimensions is the Atiyah-Hitchin metric, which is smooth everywhere and asymptotes to the negative mass Taub-NUT metric \cite{Seiberg:1996nz,Sen:1997pr}.}

At scales $r \gg r_0$ the pure $O4^-$ solution makes perfect sense however. In particular it will accurately determine the long range forces on probe particles exerted by the orientifold plane.

Although we focused on a particular example here, it is clear that similar considerations hold in general.

\subsection{Angular momentum}

From (\ref{AngularMomentum}) we see that for an orientihole system consisting of a charge $\Gamma_0$ at the origin and a charge $\Gamma_1$ together with its image $\Gamma_{-1}$, the classical field angular momentum is
\begin{equation} \label{Jor}
 \vec J = \frac{1}{4} \biggl( \langle \Gamma_1 , \Gamma_{-1} \rangle
 + \langle \Gamma_1, \Gamma_{0} \rangle + \langle \Gamma_0, \Gamma_{-1} \rangle \biggr) \vec{u} = \frac{1}{2} \biggl( \frac{1}{2} \langle \Gamma_1 , \Gamma_1' \rangle
 + \langle \Gamma_1, \Gamma_{0} \rangle \biggr) \vec{u}
\end{equation}
where $\vec{u}$ is the unit vector ${\vec x}_1/|{\vec x}_1|$. The extra overall factor $1/2$ compared to the unorientifolded case arises because of the $\IZ_2$ identification of space, which means we should integrate the angular momentum density over only half of $\IR^3$. For the last step we used (\ref{symplprodrule}).

For arbitrary orientihole configurations, we get, again using (\ref{symplprodrule}):
\begin{align}
\vec J & = \frac{1}{2} \biggl(\sum_{0<s<t}\langle \Gamma_s , \Gamma_t \rangle\frac{\vec x_{s}-\vec x_{t}}{|\vec x_{s}-\vec x_{t}|} + \langle \Gamma_s , \Gamma_t' \rangle \frac{\vec x_{s}+\vec x_{t}}{|\vec x_{s}+\vec x_{t}|}
+\sum_{s>0}  \langle \Gamma_{s},\frac{1}{2}\Gamma_{-s}  + \Gamma_{0} \rangle\frac{\vec x_{s}}{|\vec x_{s}|} \biggr) .
\end{align}

\subsection{Marginal stability decays and flow trees} \label{sec:MS}

Just as in the usual $\CN=2$ case \cite{Denef:2000nb}, multicentered BPS solutions can decay when crossing certain real codimension 1 walls in moduli space, called walls of marginal stability. Let us consider first the simplest case, a system of charges $\Gamma_0$, $\Gamma_{\pm 1}$. Then the integrability conditions (\ref{IntegrabilityConditions}) boil down to
\begin{equation} \label{eqsep}
 \frac{I(\Gamma_1,\Gamma_0)}{|{\vec x}_1|} =  \left\{
 \begin{array}{l}
  2\, {\rm Im} \, Z(\Gamma_1,t)   \quad (\epsilon=0) \\
  2\, {\rm Re} \, Z(\Gamma_1,t)   \quad (\epsilon=1)
 \end{array}
 \right.
\end{equation}
where $Z$ in the large volume approximation is given by (\ref{Zexpr}) and
\begin{equation} \label{Idef}
 I(\Gamma_1,\Gamma_0) :=  \frac{1}{2} \langle \Gamma_1 , \Gamma_1' \rangle
 + \langle \Gamma_1, \Gamma_{0} \rangle \, ,
\end{equation}
which we recognize as twice the classical field angular momentum (\ref{Jor}). Therefore, for the bound state to exist, we need
\begin{equation} \label{stabcond}
 I(\Gamma_1,\Gamma_0) \, {\rm Im} \, Z_1 > 0   \quad (\epsilon=0) \, , \qquad  I(\Gamma_1,\Gamma_0) \, {\rm Re} \, Z_1 > 0   \quad (\epsilon=1) \, .
\end{equation}
This condition determines on which side of a wall of marginal stability the bound state exists. On the wall
of marginal stability all constituents preserve the same supersymmetry, i.e.\ the phase of the central charge of $\Gamma_1$ aligns with $\alpha_\infty$ as given in (\ref{alphaval}). Specifically, $Z_1$ must be \emph{positive} real in the O4/O0 case $\epsilon=0$, and \emph{negative} imaginary in the O6/O2 case $\epsilon=1$. On such a wall, the mass of the BPS bound state equals the sum of the masses of the constituents. One should keep in mind here that
the mass of the charge at the origin $\Gamma_0$, as given by (\ref{massformula}), can be negative, namely when the phase of its central charge \emph{anti}-aligns with $\alpha_\infty$. This is the case when the origin contains only the orientifold plane, or the orientifold plane with too little additional D-branes on top of it to make the mass positive.

As a toy example, consider again the one modulus, $\epsilon=0$ case. Assume
\begin{eqnarray}
 \Gamma_1 &=& e^{a D} \, , %= 1 + a D + \frac{a^2}{2} D^2 + \frac{a^3}{6} \kappa dV
 %\Gamma_0 &=& -4D \, ,
\end{eqnarray}
where $D^3>0$ and $a$ is an arbitrary positive integer. This represents the charge of a D6-brane carrying $a$ units of flux (ignoring curvature induced charges). Then for $t=b+iv=i y D$, using (\ref{intrsympl}) and (\ref{Omvalinf}),
\begin{equation}
 Z_1 = \langle \Gamma_1 , \Omega(t) \rangle = -\int_X e^{a D} \wedge \frac{e^{-i y D}}{\sqrt{\frac{4}{3} y^3}} = \frac{D^3 (iy-a)^3}{\sqrt{\frac{4}{3} y^3}} \, .
\end{equation}
This becomes real and positive when $-a+i y$ is positively proportional to $e^{2 \pi i/3}$, i.e.\ at
\begin{equation}
 y = y_{\rm ms} = \sqrt{3} a \, .
\end{equation}
This is the ``wall'' of marginal stability --- in this case, since the moduli space is one dimensional, it is a point.

Let us assume furthermore that at the origin we have an orientifold plane with charge
\begin{equation}
 \Gamma_0 = - 4 D \, .
\end{equation}
This is modeled on (\ref{GamO4}); for simplicity we again ignore curvature induced charges. Then
\begin{equation}
 I(\Gamma_1,\Gamma_0) = \frac{1}{2} \langle e^{aD} , -e^{-aD} \rangle
 + \langle e^{aD}, -4D \rangle = 2 D^3 a^2(1-\frac{a}{3}) \, .
\end{equation}
The stability condition (\ref{stabcond}) becomes
\begin{equation}
 (a-3) (y-\sqrt{3} a) > 0 \, .
\end{equation}
%In the large $a$ regime $y_{\rm ms}$ is safely in the large volume region of moduli space, so we can trust the large volume approximation we made, and in this case the stability condition is $y>y_{\rm ms}=\sqrt{3} a$.

Similar considerations can be made for decays involving clusters of multiple centers, as in the unorientifolded case; see \cite{Denef:2007vg} for a detailed discussion.

\begin{figure}[h]
\begin{center}
\includegraphics[height=12cm]{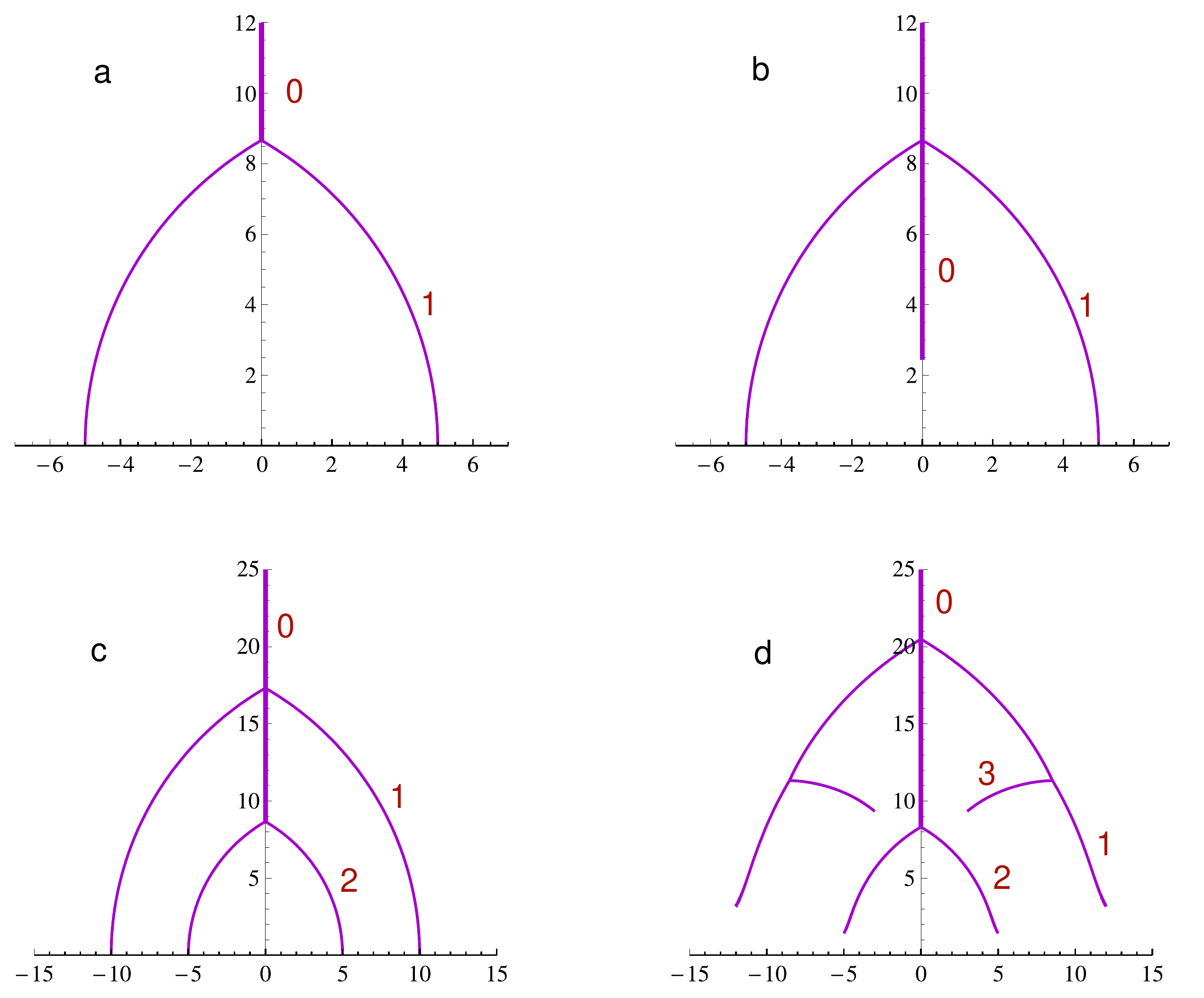}
\end{center}
\caption{Four $\epsilon=0$ orientihole flow trees. The $x$ and $y$ axes parametrize the modulus $t=(x+iy)D$. The initial point of the flow in each case is $t=30 \, i$ (or some other sufficiently large imaginary value). The constituent charges are in {\it (a)}: $\Gamma_1=e^{5D}$, $\Gamma_0=-4D$, {\it (b)}: $\Gamma_1=e^{5D}$, $\Gamma_0=10 D - 10 D^3$, {\it (c)}: $\Gamma_1=e^{10 D}$, $\Gamma_2=e^{5 D}$, $\Gamma_0=-4D$, {\it (d)}: $\Gamma_1=e^{12 D}(1-5D^2)$, $\Gamma_2=e^{5 D}(1-D^2)$, $\Gamma_3=D+3D^2-10 D^3$, $\Gamma_0=-4D$. All flows run downwards except the $\Gamma_0$ flow in {\it (a), (c), (d)} which runs up starting from the lowest vertex.
 \label{splitflows}}
\end{figure}

As in the unorientifolded case, solving the integrability condition is not sufficient for existence. In addition, it is more obscure now what ``existence'' actually means, since for example the solution corresponding to just the orientifold plane at the origin is in fact singular and of the kind we would normally reject, but now want to retain. We will consider a configuration to exist physically if we can adiabatically assemble it from constituents we know exist (e.g.\ regular single centered black holes, the pure O-plane, \ldots), by dialing the moduli at infinity through walls of marginal stability from unstable to stable side, or through walls of threshold stability. The constituents themselves could be multicentered configurations too --- the existence of those at the relevant points in moduli space is in turn determined by the ability to assemble those configurations from constituents with established existence, and so on. The justification for this working definition is that we expect BPS states to disappear only when crossing walls of marginal stability from the stable to the unstable side, so never during the assembly process, and that in all examples of regular multicentered solutions where this has been checked, the configuration can be assembled in this way.

In the unorientifolded case, a canonical prescription for dialing the moduli to produce such an assembly process is given by \emph{attractor flow trees}, which were briefly reviewed in section \ref{sec:revN2}, and more extensively in section 3.2 of \cite{Denef:2007vg}. Up to some differences which we discuss below, we can do the same in the orientifolded case. Some one modulus examples are shown in fig.\ \ref{splitflows}, with case {\it (a)} corresponding to the toy example discussed above, with $a=5$. The $\IZ_2$ orientifold symmetry is clearly manifested in the flow trees. The assembly process is dictated by following the tree in the direction from leaves to root; for example in case {\it (a)}  we start with a fluxed D6-brane (which we know exists) and put it together with its orientifold image at the split point, after which we move upwards.

The main difference with unorientifolded flow trees is that tertiary vertices occur generically; they correspond to splits $\Gamma \to \Gamma_1 + \Gamma_0 + \Gamma_1'$. In the unorientifolded case, one can always make a small perturbation of the initial point to split tertiary vertices in two binary vertices, but because of the orientifold projection constraints this is not possible in the case at hand. (In a sense, of course, these tertiary splits are actually binary splits, since $\Gamma_1$ and its image $\Gamma_1'$ are physically identified.) Binary vertices (with their orientifold images) are also still possible, as illustrated in fig.\ \ref{splitflows} {\it (d)}. They correspond to splits $\Gamma_{12} \to \Gamma_1 + \Gamma_2$.

Another difference with the unorientifolded case is that, as discussed earlier, \emph{inverted} attractor flows can occur too, namely for an invariant charge whose phase is opposite to $\alpha_\infty$, corresponding to a negative mass object. This is the case for the ``0'' branch in in fig.\ \ref{splitflows} {\it (a)}, {\it (c)} and {\it (d)}. Recall our formulae for single flow solutions already take this into account, so no additional inversion by hand is necessary when using these formulae to build the tree.

A more systematic analysis of flow trees for orientiholes and a precise formulation of the analog of the ``split flow conjecture'' of \cite{Denef:2007vg} will be left for future work.

Finally, when a wall of marginal stability is crossed, the number of BPS states can jump. We return to this in section \ref{sec:counting}.

\subsection{Dirac quantization}

\begin{figure}[h]
\begin{center}
\includegraphics[height=1cm]{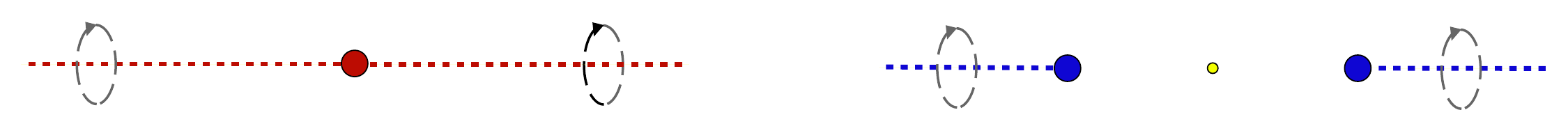}
\end{center}
\caption{Left: Dirac strings for charge $\Gamma_0$ at origin. Right: Dirac strings for charge pairs $\Gamma_{\pm s}$. \label{Dirac}}
\end{figure}

If we insist on using a single gauge connection 1-form rather than work with different connection 1-forms in different patches related by gauge transformations, the gauge fields will have Dirac string singularities as soon as there is magnetic charge present. These strings should not be physical, i.e.\ no experiment should be able to locally measure a Dirac string singularity. Since electric charges moved along a closed loop pick up a measurable wave function phase, this requirement implies charge must be quantized in such a way that the phase factor which is picked up when moving along an infinitesimally small loop around a Dirac string equals 1 and is hence undetectable.

The orientifold projection conditions imply that Dirac strings should be $\IZ_2$ (anti-)symmetric about the origin. For charges away from the origin, this can always be arranged by taking the Dirac strings attached to the charges to lie along radially outward lines in the $\IR^3$ coordinate space, as shown in fig.\ \ref{Dirac}. The Dirac quantization constraints then remain manifestly unchanged compared to the unorientifolded case, so no further quantization conditions should be imposed on charges away from the origin beyond those which existed in the parent ${\cal N}=2$ theory.

On the other hand, for charges localized \emph{at} the origin, the requirement of $\IZ_2$ symmetry implies a \emph{doubling} of charge quanta. This is because a minimal magnetic charge quantum (with respect to an arbitrary $U(1)$ subgroup) of the unorientifolded theory corresponds to a ``primitive'' Dirac string as seen by a suitably electrically charged probe particle, which cannot be split in two strings without violating the requirement of local undetectability. But such a primitive string violates the required $\IZ_2$ symmetry. For two units of magnetic charge, this problem is avoided because in this case the string can be split in a $\IZ_2$ symmetric fashion, as shown in fig.\ \ref{Dirac}. The charge doubling holds for both electric and magnetic charges, as for electric charges at the origin we can similarly use magnetic monopole probe charges localized away from the origin.

Thus we conclude that if in the covering space the charge lattice is $L \simeq \IZ^{2n_V+2}$, then quantum consistency after orientifolding requires
\begin{equation} \label{DiracQuant}
 \Gamma_0 \in 2 L \, , \qquad \Gamma_s \in L  \quad (s \neq 0) \, .
\end{equation}
In the quotient space $\IR^3/\IZ_2$ this is equivalent to the statement that electromagnetic field flux through closed surfaces is integrally quantized in the standard way: the relative factor of 2 above then simply corresponds to the fact that \emph{half} a sphere in $\IR^3$ with center at the origin is already a closed surface in $\IR^3/\IZ_2$, while charges away from the origin need a full sphere around them to be enclosed.

%In the unorientifolded theory, Dirac quantization is equivalent to half-integrality of the field angular momentum sourced by any pair of charges. In the orientifolded theory, as (\ref{Jor}) shows, Dirac quantization is slightly stronger, since non-doubled charges $\Gamma_0$ would still give half-integral angular momenta ($\langle \Gamma_s,\Gamma_{-s} \rangle$ is always even for integral charges, as can be seen using (\ref{mapsms})).

We give an alternative derivation based on the gauge bundle description of magnetic monopoles in appendix \ref{app:gaugebundle}.

\subsection{$\IZ_2$ torsion charges} \label{sec:torsion}

\begin{figure}[h]
\begin{center}
\includegraphics[height=6cm]{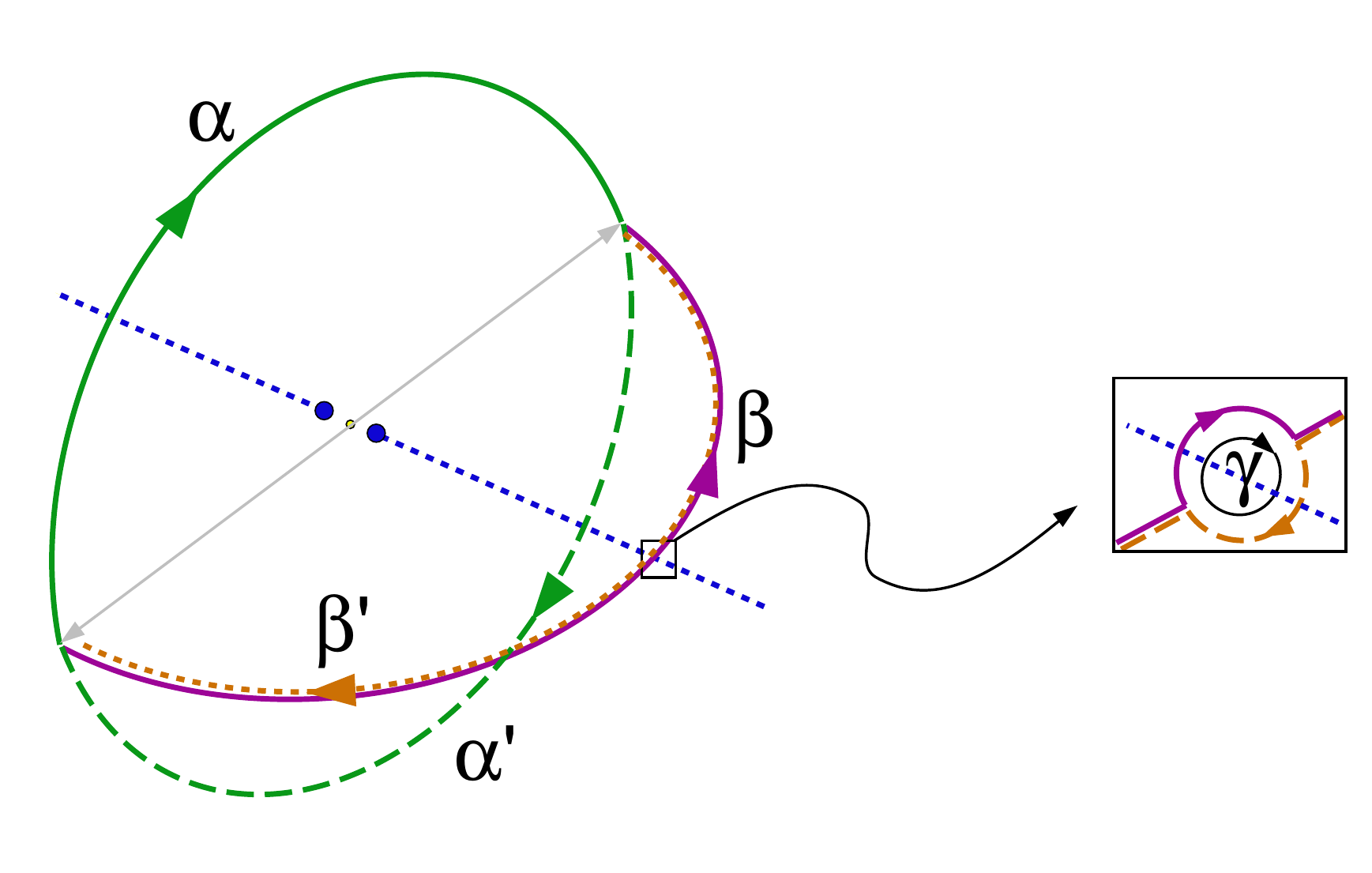}
\end{center}
\caption{An electric probe transported along $\alpha$ picks up an Aharonov-Bohm phase, measuring a magnetic $\IZ_2$ torsion charge. See text for more details. \label{torsion}}
\end{figure}

Although (\ref{orinvQ}) indicates that certain net D-brane charges have to vanish, there can still be conserved $\IZ_2$ torsion charges associated to these. For example there can be no net D6 charge in an O4/O0 type orientifold, but an orientihole configuration with $\sum_{s>0} P^0_s$ odd has a nonvanishing torsion charge, which can be measured at spatial infinity by an Aharonov-Bohm type experiment. To see this, let us first consider the O4/O0 case ($\epsilon=0$) with a charge $\Gamma_1$ and its image $\Gamma_{-1}$, where $\Gamma_1$ has one unit of D6-charge, so $\Gamma_{-1}$ has minus one unit of D6-charge. We take the corresponding Dirac strings as before to go radially out to infinity.\footnote{Alternatively we can take the Dirac string to be stretched between the two charges, not running to infinity. This case will be further discussed below. \label{altern}} Let $\alpha$ be a semi-circle of very large radius $R$ in the plane equidistant from $\Gamma_1$ and $\Gamma_{-1}$ centered around the origin, as in fig.\ \ref{torsion}. The endpoints of this semi-circle are identified by the $\IZ_2$ transformation $\vec x \to -\vec x$, so in the physical, quotiented space, $\alpha$ is a closed trajectory. A D0 probe particle transported along $\alpha$ will pick up a phase $e^{i \Phi(\alpha)}$, measurable in an Aharonov-Bohm type experiment. We claim this phase is $-1$ in the limit $R \to \infty$.

To prove this without writing out explicit expressions for the gauge fields, we proceed as follows. Let $\alpha'$ be the $\IZ_2$ image of $\alpha$. Then the $\IZ_2$ symmetry implies $\Phi(\alpha')=\Phi(\alpha)$.\footnote{This can be read off directly from the projection conditions, but to eradicate all doubt, the other conceivable possibility, $\Phi(\alpha')=-\Phi(\alpha)$, is easily excluded, since $\Phi(\alpha)+\Phi(\alpha')$ is proportional to the magnetic flux through the disk bounded by $\alpha+\alpha'$, which is nonvanishing since $\Gamma_{\pm 1}$ have opposite magnetic charges.} Let $\beta$ be the curve obtained by rotating $\alpha$ keeping its endpoints fixed till right before it hits the Dirac string attached to $\Gamma_1$, and $\beta'$ the curve similarly obtained from $\alpha'$, as shown in fig.\ \ref{torsion}. The difference $\Phi(\alpha) - \Phi(\beta)$ is proportional to the magnetic flux through the quarter sphere swept out during the rotation of $C$. However since the far magnetic field is that of a magnetic dipole, it falls off as $1/r^3$ and therefore in the limit $R \to \infty$, the magnetic flux is zero and $\Phi(\beta)=\Phi(\alpha)$. Similarly $\Phi(\beta')=\Phi(\alpha')$. Thus we have
\begin{equation} \label{Phi}
 \Phi(\alpha) = \frac{\Phi(\alpha)+\Phi(\alpha')}{2} = \frac{\Phi(\beta)+\Phi(\beta')}{2} \, .
\end{equation}
Now $\beta$ and $\beta'$ are each others inverse except for the side at which they pass the Dirac string. As a result, the right hand side of (\ref{Phi}) equals half the phase picked up when circling once closely around the Dirac string, i.e.\
\begin{equation}
 \Phi(\alpha) = \frac{1}{2} \Phi(\gamma) \, ,
\end{equation}
where $\gamma$ is an infinitesimally small loop around the Dirac string. For a minimal magnetic quantum, i.e.\ unit D6-charge, we have by definition $\Phi(\gamma) = 2 \pi$, hence $\Phi(\alpha)=\pi$ and $e^{i \Phi(\alpha)}=-1$, as claimed.

More generally, for general $\epsilon=0$ orientihole configurations, one similarly shows that in the limit $R \to \infty$
\begin{equation} \label{torsionformula}
 e^{i \Phi(\alpha)} = (-1)^{\sum_{s>0} P_s^0}.
\end{equation}
This defines a topological $\IZ_2$ torsion charge. This is a conserved charge in the sense that even if the whole system is thrown into a big (not necessarily supersymmetric) black hole, the charge remains measurable by the same type of experiment. The $\IZ_2$ nature of the charge is compatible with the fact that the orientifold $\IZ_2$ quotient of the sphere at infinity is $S^2/\IZ_2=\IRP^2$, and $H_1(\IRP^2,\IZ)=\IZ_2$, with generator given by the path $\alpha$. Furthermore the fact that an even number of D6-anti-D6 pairs has zero $\IZ_2$ charge is in accord with the fact that in this case, D6 and anti-D6 branes of \emph{different} pairs can annihilate each other completely.

One could also add this D6 torsion charge to the O4-plane at the origin itself. One can think of this as being produced by squeezing together a D6 and an anti-D6 at the origin. In this case there will be a D6 dipole Dirac string coming out of the origin in two opposite directions. If we had chosen the Dirac string in our original configuration to be stretched between the two charges as described in footnote \ref{altern}, this would have corresponded to a situation with nontrivial $\IZ_2$ charge at the origin, plus nontrivial $\IZ_2$ charge from the charge pair, leading to a vanishing total $\IZ_2$ charge.

Finally, one can clearly extend these considerations to all other charges projected out by the orientifold symmetry, i.e.\ $P^0$, $P^-$ and $Q_+$  for $\epsilon=0$ and $P^+$, $Q_-$ and $Q_0$ for $\epsilon=1$, all of which will support $\IZ_2$ torsion charges.

In a compact setting (i.e.\ compactifying $\IR^3$ to $T^3$), or on the T-dual side with space-filling D-branes, all charges, including the $\IZ_2$ torsion charges just described, must vanish. In the model building literature, this requirement is often referred to somewhat loosely as ``K-theory tadpole cancelation'', and in practice the $\IZ_2$ torsion charges are usually only indirectly detected \cite{Uranga:2000xp} by the presence of subtle anomalies on probe branes such as the $SU(2)$ anomaly described in \cite{Witten:1982fp}. The existence of these $\IZ_2$ tadpoles in the D9-D9$'$ system was crucially used in \cite{Collinucci:2008pf} to find agreement between F-theory and perturbative type IIB descriptions of D7-branes. Here we see that in the orientihole picture, the $\IZ_2$ charges get an elementary physical interpretation. Of course, this is only a tiny fraction of what K-theory has to say about orientifolds \cite{mooretexas,hori}.

\section{Counting BPS states} \label{sec:counting}

\subsection{BPS index}

In an $\CN=2$ theory, the proper index counting the number of BPS states of charge $\Gamma$ is the second helicity supertrace \cite{Cecotti:1992qh,Kiritsis:1997gu}:
\begin{equation}
\Omega(\Gamma,t):=-\frac{1}{2}\mathrm{Tr}_{{}_{{\cal H}(\Gamma,t)}} (-)^{2J_3}(2J_3)^2
\end{equation}
where $J_3$ is the $3$-component of the angular momentum.
Under the little group $SU(2)$ of the super-Poincar\'e symmetry, massive BPS multiplets decompose in representations of the form $[j']\otimes ([\sm{\frac{1}{2}}]+2[0])$. The half-hypermultiplet factor $([\sm{\frac{1}{2}}]+2[0])$
arises from the quantization of the fermionic degrees of freedom associated to the center of mass in $\mathbb{R}^3$ (alternatively, from the broken supersymmetries) and $j'$ is the reduced angular momentum associated to the internal degrees of freedom.
In particular, $j'=0$ corresponds to a hypermultiplet and $j'=\sm{\frac{1}{2}}$ to a vector multiplet.
If we express the second helicity supertrace in term of the reduced angular momentum, it becomes an ordinary Witten index:
\begin{equation}
\Omega(\Gamma,t)=\mathrm{Tr}_{{}_{{\cal H'}(\Gamma,t)}} (-)^{2J_3'},
\end{equation}
where ${\cal H'}(\Gamma,t)$ is obtained from ${\cal H}(\Gamma,t)$ by stripping off the overall factor $([\sm{\frac{1}{2}}]+2[0])$.
In the orientifolded theories we have been considering, we will use the same index, but trace only over the orientifold invariant subspace ${\cal H'}_{\rm inv}(\Gamma,t)$:
\begin{equation} \label{WittenIndex}
 \Omega_{\rm inv}(\Gamma,t) = \mathrm{Tr}_{{}_{{\cal H'_{\rm inv}}(\Gamma,t)}} (-)^{2J_3'} \, .
\end{equation}

\subsection{Wall crossing formula}

As discussed in section \ref{sec:MS}, the simplest possible decay upon crossing a wall of marginal stability is of the form $\Gamma \to \Gamma_1 + \Gamma_0 + \Gamma_1'$. Here we take $\Gamma_1$ and $\Gamma_0$ to be primitive (not an integral multiple of another charge). In the supergravity description, these charges can be realized either as single or as multicentered bound states.

This decay will lead to a jump $\Delta \Omega_{\rm inv}(\Gamma,t)$ of the BPS index. To compute this, we
follow the reasoning of \cite{Denef:2002ru,Denef:2007vg}. When $t \to t_{\rm ms}$, the part of the BPS Hilbert space which is decaying factorizes as
\begin{equation}
 \Delta {\cal H'}_{\rm inv}(\Gamma,t_{\rm ms}) = [j] \otimes {\cal H'}_{\rm inv}(\Gamma_0,t_{\rm ms}) \otimes {\cal H'}(\Gamma_1,t_{\rm ms}) \, .
\end{equation}
Here $j$ is the angular momentum quantum number of the BPS ground state of the supersymmetric quantum mechanical system obtained by replacing the $\Gamma_1$ system by a point charge without internal degrees of freedom. Assuming the effect of the spin-magnetic coupling is, as in \cite{Denef:2002ru}, to reduce the classical angular momentum by $-1/2$ in the quantum ground state, and using (\ref{Jor}), we have $j=\frac{I(\Gamma_1,\Gamma_0)-1}{2}$, where $I(\Gamma_1,\Gamma_0)$ was defined in (\ref{Idef}):
\begin{equation} \label{Idefrep}
 I(\Gamma_1,\Gamma_0) =  \frac{1}{2} \langle \Gamma_1 , \Gamma_1' \rangle
 + \langle \Gamma_1, \Gamma_{0} \rangle \, .
\end{equation}
Notice that because of the orientifold identification of $\Gamma_1$ and $\Gamma_1'$, we have only a single factor ${\cal H}'(\Gamma_1)$ and not ${\cal H}'(\Gamma_1) \otimes {\cal H}'(\Gamma_1')$, as we would if we were considering the unorientifolded theory. On the other hand we do not project on the invariant subspace for this charge, whereas we do for $\Gamma_0$.

Thus we are led to the following wall crossing formula:
\begin{equation} \label{wcf}
 \Delta \Omega_{\rm inv} = (-1)^{I(\Gamma_1,\Gamma_0)-1} \, |I(\Gamma_1,\Gamma_0)| \, \Omega(\Gamma_1,t_{\rm ms}) \, \Omega_{\rm inv}(\Gamma_0,t_{\rm ms}) \, .
\end{equation}
This is formally identical to the primitive charge wall crossing formula of \cite{Denef:2007vg} for a decay $\Gamma \to \Gamma_1 + \Gamma_0$, after replacing $\langle \Gamma_1,\Gamma_0 \rangle$ appearing there by $I(\Gamma_1,\Gamma_0)$, and replacing $\Omega(\Gamma_0)$ by $\Omega_{\rm inv}(\Gamma_0)$.

We leave the generalization to wall crossing formulae involving nonprimitive charges, analogous to the formulae given in \cite{Denef:2007vg,KS,Gaiotto:2008cd}, to future investigations.

If the constituent branes of a particular configuration are rigid, as is the case for single D6-branes carrying only smooth $U(1)$ fluxes, repeated use of the wall crossing formula (or the ideas leading to it) is sufficient to compute the total index. As a simple illustration, consider again the basic example of section \ref{sec:MS}: a charge $\Gamma_1 = e^{a D}$ plus the orientifold plane charge $\Gamma_0 = - 4 D$. Then $\Omega(\Gamma_1)=1$ and $I(\Gamma_1,\Gamma_0)=2 D^3 a^2(1-\frac{a}{3})$, so the total index of the configuration at $y>y_{\rm ms} = \sqrt{3} a$ is, assuming $a>3$,\footnote{We also assume $D^3$ is a multiple of 3 such that the index is integral for all $a$, as is necessary for consistency. Recall this is a toy example; in actual examples, things will conspire such that indices are indeed always integral.}
\begin{equation}
 \Omega_{\rm inv} = 2 D^3 a^2(\frac{a}{3}-1) \, .
\end{equation}

\subsection{Microscopic interpretation} \label{sec:microint}

The wall crossing formula (\ref{wcf}) also has a simple microscopical (D-brane) interpretation, in close analogy to the unorientifolded case \cite{Denef:2002ru,Denef:2007vg}. For simplicity we let $\Gamma_0$ consist of just the orientifold plane, so we are considering a bound state of a D-brane of charge $\Gamma_1$ with its orientifold image.
In the weak string coupling limit $g_s \to 0$, the bound state will collapse to the origin of space \cite{Denef:2002ru} and the index (\ref{WittenIndex}) can be thought of as the\footnote{When the moduli space is singular, as is often the case, there are several inequivalent notions of Euler characteristic. We will not try to determine in general which notion is the physically relevant one, although in the example we will study below, it appears to be the topological one.} Euler characteristic of the moduli space $\CM_{\rm inv}(\Gamma,t)$ of orientifold invariant supersymmetric D-brane configurations with total charge $\Gamma$. More precisely, properly taking into account the sign factors and the identification of spatial spin with Lefshetz spin (as explained e.g.\ in \cite{Denef:2002ru,Denef:2007vg}):
\begin{equation} \label{Omchi}
 \Omega_{\rm inv} = (-1)^{\dim_\IC \CM_{\rm inv}} \, \chi(\CM_{\rm inv}) \, .
\end{equation}
The moduli space can have several disconnected components. The component decaying at the wall of marginal stability is expected \cite{Denef:2007vg} to have the structure of a $\ICP^{I-1}$-fibration
\begin{equation}
 \ICP^{I-1} \to \Delta \CM_{\rm inv}(\Gamma,t) \to \CM(\Gamma_1,t_{\rm ms})
\end{equation}
where $I$ is the index counting orientifold invariant massless open string modes stretched between the constituent branes. This index was computed in \cite{Brunner:2003zm,Brunner:2004zd,Collinucci:2008pf}. In particular, the results of \cite{Collinucci:2008pf} imply (taking into account the factor $1/8$ difference in charge between an O4 and an O7 wrapping the same internal cycle) that this index $I$ is nothing but $I(\Gamma_1,\Gamma_0)$ defined in (\ref{Idefrep}), with $\Gamma_0$ the orientifold plane charge. Using the fact that the Euler characteristic of the moduli space of a nondegenerate fibration equals the product of the Euler characteristics of base and fiber, we thus reproduce exactly the wall crossing formula (\ref{wcf}) in this particular case.

A corollary is the following general rule:

\begin{center}
 open string index for pair of branes \\
 = \\
 twice the classical field angular momentum generated by the pair
\end{center}

\noindent This holds for a brane image brane pair as well as for two distinct branes.

\subsection{Relation to topological string partition function: a conjecture}

In \cite{Ooguri:2004zv}, Ooguri, Strominger and Vafa (OSV) proposed a conjecture extending the Bekenstein-Hawking entropy formula for BPS black holes to all orders in an inverse charge expansion:
\begin{equation} \label{OSVintform}
 \Omega(P,Q) \sim \int d\phi \, e^{2 \pi \phi^\Lambda Q_\Lambda
 } \, |{\cal Z}_{\rm top}|^2,
\end{equation}
where ${\cal Z}_{\rm top}(g,t)$ is the topological string partition
function in which the following substitutions are made
\begin{equation} \label{osvsubst}
  g =
  \frac{4\pi}{2 \phi^0 + i \, P^0 }, \qquad
  t^A = \frac{2 \phi^A + i \, P^A}{2 \phi^0 + i \, P^0}.
\end{equation}
More explicitly
\begin{eqnarray}
 {\cal Z}_{\rm top}(g,t) &=& \exp \biggl( -\frac{(2\pi i)^3}{6 g^2} t^3
 - \frac{2 \pi i}{24} c_2 t + \sum_{\beta,h}
 N_{h,\beta} \, (-)^{h-1}
 g^{2h-2} \, e^{2 \pi i \beta_A t^A} \biggr),
\end{eqnarray}
where $c_2$ is the second Chern class of $X$ and $N_{h,\beta}$ are the Gromov-Witten invariants counting holomorphic
maps of genus $h$ into the class $\beta \in H_2(X,\IZ)$.

The main motivation for the OSV conjecture given by \cite{Ooguri:2004zv} was the observation that the saddle point evaluation of the integral (\ref{OSVintform}) exactly reproduces the ($R^2$ corrected) attractor equations found in \cite{LopesCardoso:1998wt,LopesCardoso:1999cv,LopesCardoso:1999xn}, with saddle point value equal to $e^{S_{\rm BHW}(P,Q)}$, where $S_{\rm BHW}$ is the Bekenstein-Hawking-Wald entropy \cite{LopesCardoso:1998wt,LopesCardoso:1999cv,LopesCardoso:1999xn} (which takes into account $R^2$ corrections determined by topological string amplitudes).

It is difficult to make the OSV conjecture precise. Un fact it has be shown to be false in the large charge limit when $\Omega$ is taken to be the index at ${\rm Im} \, t = \infty$ \cite{Denef:2007vg}. This leaves open the possibility that the conjecture is true when $\Omega$ is taken to be the index at the attractor point of the charge under consideration. We will not dwell on these difficulties, but just formulate an analogous conjecture for orientiholes, at the same level of precision as \cite{Ooguri:2004zv}. We conjecture:
\begin{equation} \label{DEPintform}
 \Omega_{\rm inv}(P,Q) \sim \int d\phi \, e^{\pi \phi^\Lambda Q_\Lambda
 } \, {\cal Z}_{\rm top},
\end{equation}
where again the substitutions (\ref{osvsubst}) are understood, and $P$, $Q$ and $\phi$ are restricted to the orientifold invariant sector. Notice the factor of $2$ difference with (\ref{OSVintform}) in the exponential, and the absence of the square --- the relation to the topological string partition function is now linear!

We will give arguments in favor of this conjecture below, but for concreteness let us first write out things somewhat more explicitly for $\epsilon=0$. In this case we retain $P^{A+}$, $Q_{A-}$ and $Q_0$ charges, $\phi^{A-}$ and $\phi^0$ potentials, and the substitutions are
\begin{equation}
 g = \frac{2\pi}{\phi^0} \, , \qquad t^{A+} = i \frac{P^{A+}}{2 \phi^0} \, , \qquad t^{A-} = \frac{\phi^{A-}}{\phi^0} \, .
\end{equation}
This gives
$$
 \CZ_{\rm top} = \exp \left(
 \frac{\pi \left( (P^+)^3 + c_2 P^+ \right)}{24 \, \phi^0}  - \frac{\pi P^+ (\phi^-)^2}{2 \, \phi^0}
 + \sum_{\beta,h} N_{h,\beta} (-)^{h-1} \left( \frac{2 \pi}{\phi^0} \right)^{2h-2} \,
 e^{- \frac{\pi \beta_{+} P^{+}}{\phi^0}}  \, \cos \left( \frac{2 \pi \beta_{-} \phi^{-}}{\phi^0} \right)
 \right)
$$
To get the cosine we used the invariance of $N_{h,\beta}$ under the orientifold transformation. Note this expression is manifestly real, as it should for (\ref{DEPintform}) to make sense.

The $\epsilon=1$ case can be worked out similarly, and again everything is real.

To motivate this conjecture along the lines of \cite{Ooguri:2004zv}, first observe that because of its reality, the integrand of (\ref{DEPintform}) is the square root of the one in (\ref{OSVintform}), restricted to the orientifold invariant locus. Since this amounts to an overall factor $1/2$ in the exponential, the saddle point equations will therefore still coincide with the attractor equations for the orientihole of charge $(P,Q)$, and the saddle point value of the integral will be
\begin{equation}
 e^{\frac{1}{2} S_{\rm BHW}(P,Q)} \, ,
\end{equation}
where $S_{\rm BHW}$ is again the Bekenstein-Hawking-Wald entropy. The extra factor $1/2$ is precisely what we should have for an orientihole.

Another (sketchy) argument is this: In the proof of \cite{Denef:2007vg} of a version of the OSV conjecture, the $\CZ_{\rm top} \CZ_{\rm top}^*$ structure is ultimately due to the factorization of the extreme polar part of the D4 partition function in a D6 partition function and an anti-D6 partition function. This in turn is due to the fact that extreme polar BPS states split as bound states of D6- and anti-D6-branes. The D6 partition function is identified with the Donaldson-Thomas partition function, which in turn is identified with the topological string partition function, leading to $\CZ_{\rm D4} \sim \CZ_{\rm top} \CZ_{\rm top}^*$ and (\ref{OSVintform}). Now, one could imagine constructing a similar derivation in the orientifolded theory. Again extreme polar D4 states will split in bound states of D6- and anti-D6-branes. The difference is that now the D6 and anti-D6 will be \emph{identified} by the orientifold action, as illustrated by the examples considered earlier. Thus, we expect only a \emph{single} factor $\CZ_{\rm top}$, leading to (\ref{DEPintform}).

We leave a more careful investigation of these arguments to future work.

%For $\epsilon=1$, we retain $P^0$, $P^{A-}$ and $Q_{A+}$ charges, $\phi^{A+}$ potentials, and the substitutions are
%\begin{equation}
% g = -\frac{4 \pi i}{P^0} \, \qquad t^{A+} = i \frac{-2 \phi^{A+}}{P^0} \, , \qquad t^{A-} = \frac{p^{A-}}{P^0} \, .
%\end{equation}

\section{Application to counting D7 vacua} \label{sec:countingD7}

We can now apply the orientihole technology we developed to the problem of counting D7 vacua in IIB orientifolds, as outlined in section \ref{sec:vactoOr}. By ``counting'' D7 vacua we mean, as in the vacuum statistics program laid out in \cite{Douglas:2003um}, computing the Witten index, i.e.\ the Euler characteristic of the classical moduli space of supersymmetric D7-configurations, in the absence of background bulk fluxes. This includes bound states with lower dimensional branes, realized as sheaves in the large volume limit. In general there will be many disconnected components of this moduli space. Some of these will be points, corresponding to sectors with all D-brane moduli fixed classically. Sectors with residual moduli could have their moduli lifted by quantum effects, perhaps after supersymmetry breaking. If so, by Morse theory, the number of critical points of the induced potential, counted with signs, will be bounded below by the Euler characteristic, and the latter can often be used as a good generic estimate for the former. This justifies to some extent our working definition of ``counting'' vacua. We should note however that more subtle effects may be missed by this definition, such as the existence of $N$ quantum vacua of pure $SU(N)$ 4d $\CN=1$ super Yang-Mills realized by some rigid brane system. It will suffice though for many purposes.

To be concrete, we consider again the Calabi-Yau described in section \ref{sec:simple}, the degree eight hypersurface in $\ICP^4_{4,1,1,1,1}$:
\begin{equation} \label{CYeq2}
 z_0^2 = h(z_1,z_2,z_3,z_4) \, .
\end{equation}
This was the main example studied in \cite{Collinucci:2008pf}, where the O7/O3 orientifold with involution $\sigma:z_0 \to -z_0$ was analyzed from various perspectives. It was found there that even the pure D7-brane without any worldvolume flux had already a rather intricately constrained structure. In particular it was pointed out that the D7 necessarily intersects the O7-plane $z_0=0$ in double points. Similar observations were made in \cite{Braun:2008ua} from an F-theory perspective, and recently in \cite{Brunner:2008bi} from a worldsheet perspective.

\subsection{Basic D7-branes}

\subsubsection{Flow trees} \label{sec:D7flowtrees}

\begin{figure}[h]
\begin{center}
\includegraphics[height=7cm]{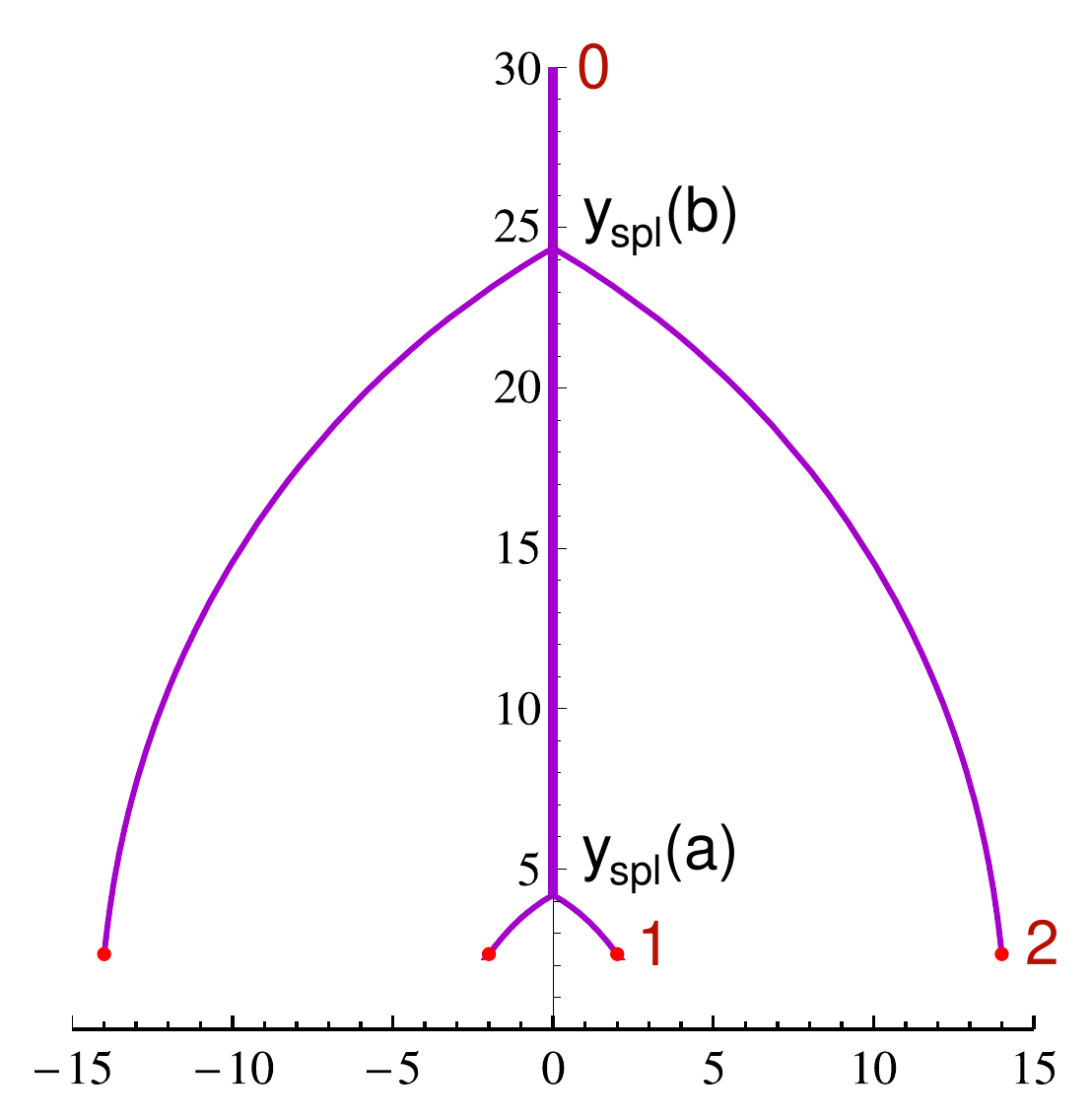}
\end{center}
\caption{Flow tree for the case $a=2$, $b=14$, with $t=(x+iy) D$. \label{fl214}}
\end{figure}

One particularly useful picture of D7-branes considered in \cite{Collinucci:2008pf} is their description as bound states of D9 and image anti-D9 branes. In the orientihole dual this maps to bound states of D6 and image anti-D6 branes. From anomaly considerations it was argued in \cite{Collinucci:2008pf} that there must always be an even number of D9-D9$'$ pairs. In the orientihole context, this can be understood as being due to the presence of a $\IZ_2$ torsion charge when only an odd number of D6-branes is present, as discussed in section \ref{sec:torsion}. The simplest consistent configurations of this kind are obtained by considering two pairs of D9-D9$'$ pairs, carrying $a$ resp.\ $b$ units of $U(1)$ flux. The orientihole dual thus consists of an O4$^-$ with charge given by (\ref{GamO4}) and the point charges
\begin{eqnarray}
 &&\Gamma_1 = e^{a D}(1+\sm{\frac{c_2}{24}}) \, , \qquad \Gamma_{-1} = -e^{-a D}(1+\sm{\frac{c_2}{24}}) \, , \label{ch1} \\
 &&\Gamma_2 = e^{b D}(1+\sm{\frac{c_2}{24}}) \, , \qquad \Gamma_{-2} = -e^{-b D}(1+\sm{\frac{c_2}{24}}) \, , \label{ch2}
\end{eqnarray}
where $c_2$ is the second Chern class of the Calabi-Yau; in the case at hand this is \cite{Collinucci:2008pf}
\begin{equation}
 c_2 = 22 \, D^2 \, .
\end{equation}
Bound states of this kind look like fig.\ \ref{orientiholes} and are described by flow trees similar to the one in fig.\ \ref{splitflows}{\it c}; the case $a=2$, $b=14$ is shown in fig.\ \ref{fl214}.

We can assume $a \leq b$ without loss of generality. An attractor flow tree analysis along the lines of section \ref{sec:MS} yields existence of a BPS configuration in a large internal volume background $t=i y D$ iff
\begin{equation}
 a \geq 2 \, , \qquad y > y_{\rm ms} = y_{\rm spl}(b) := \sqrt{3 b^2 + \frac{11}{2}} \, .
\end{equation}
The first charge to split off is $\Gamma_2$, at $y = y_{\rm spl}(b)$, and the second is $\Gamma_1$, at $y = y_{\rm spl}(a)$. The $I$-indices at the split points are
\begin{eqnarray} \label{Indvals1}
 I(\Gamma_1,\Gamma_0) &=& - \frac{(2a-1)(2a-2)(2a-3)}{6}  \, , \\
 I(\Gamma_2,\Gamma_1 + \Gamma_0 + \Gamma_1') &=&
 - \frac{(2b-1)(2b-2)(2b-3)}{6} - \frac{2 a^3}{3} - \frac{22 \, a}{3} \, . \label{Indvals2}
\end{eqnarray}
Some comments are in order:
\begin{enumerate}
 \item Strictly speaking, in the large volume approximation, none of the charges in (\ref{ch2}) has an attractor point --- the central charge in the large volume approximation has a zero at an interior point of the moduli space; for example for $\Gamma_1$ this lies at $t=t_0=(a+\frac{\sqrt{22}}{2}i)D$. However in this region of moduli space instanton corrections to the central charge are important. In principle, these corrections can be computed by mirror symmetry \cite{Candelas:1990rm}. In fact, a single D6 with flux is known to be mirror to a D3 wrapping a 3-cycle vanishing at a conifold point \cite{Brunner:1999jq}. Such D-branes do exist as BPS states. In \cite{Strominger:1995cz} it was pointed out that integrating them out is what causes the logarithmic singularity at the conifold point in moduli space. They do correspond to sensible attractor flows, terminating at the conifold point \cite{Moore:1998pn,Denef:2000nb}. So in principle we should include instanton corrections, as was done in the flow tree analysis of \cite{Collinucci:2008ht}, and then we would find an admissible attractor flow for these charges. But we do not have to do this --- all we need to know at this point is that the BPS state exists. One could object that instanton corrections might be important for establishing the existence of the split points of the flow tree as well. However, for $a \geq 2$ the ratio of the largest instanton contributions at $t=t_0$ and $t=i y_{\rm spl}(a)$ is
\begin{equation} \label{ratio}
 e^{-2 \pi ({\rm Im}\, t_{\rm spl} - {\rm Im}\, t_0)} < 0.01 \, ,
\end{equation}
so we can reasonably expect that instantons will not qualitatively affect the existence of the split point. To rigorously check this, one would need to do a full numerical analysis using the exact periods, which we will not do here.

\item The case $a=2$ is a somewhat subtle boundary case. Note that the charge $\Gamma_0^{(1)}:=\Gamma_1 + \Gamma_0 + \Gamma_{-1}=19 \, dV$, i.e.\ $-19$ units of D0-charge. Its central charge at $t=iyD$ is $Z=-\frac{19}{{\sqrt{\frac{8 y^3}{3}}}}$. This has phase opposite to $\alpha_\infty = 0$, so the attractor flow for this charge will be inverted, i.e.\ run towards \emph{increasing} $|Z|$, i.e.\ \emph{decreasing} $y$. So for any $b \geq a$, there will indeed be a flow down from $y_{\rm spl}(b)$ to $y_{\rm spl}(a)$. Note that $|I(\Gamma_0,\Gamma_1)|=1$ in this case, so by (\ref{wcf}) this configuration has index $\Omega = 1$. This fits well with $a=2$ being the boundary case.

\item For $a=1$, we have $\Gamma_0^{(1)} = -2 D + \frac{32}{3} \, dV$, which has negative central charge, so again its flow is inverted. The minimum of $|Z|$, which is now a ``repulsor'' rather than an attractor, is attained at $y=4$. Thus, because $y_{\rm spl}(b) > 4$ for all $b \geq 2$, the flow tree does not exist for $b \geq 2$. Because $y_{\rm spl}(1)<4$, it does not exist for $b=1$ either if the background $y>4$. Moreover, in any case, for $a=1$ we have $I(\Gamma_1,\Gamma_0)=0$, so no proper 2-centered bound state of $(\Gamma_1,\Gamma_0,\Gamma_{-1})$ can exist. We conclude that no BPS configurations exist with $a<2$.

\end{enumerate}

\subsubsection{Index from wall crossing}

Since we did not give a wall crossing formula for nonprimitive charges, we will assume $a \neq b$.
Then, by applying the wall crossing formula (\ref{wcf}) twice and using that the index of all constituents is 1, we immediately get that for $y>y_{\rm ms}$, the bound state index is
\begin{equation} \label{indresult}
 \Omega = (-1)^{I_1+I_2} I_1 I_2 \, , \qquad I_1 = |I(\Gamma_1,\Gamma_0)|, \quad I_2 = |I(\Gamma_2,\Gamma_1+\Gamma_0+\Gamma_1')| \, .
\end{equation}
More explicitly, $I_1$ and $I_2$ can be read off from (\ref{Indvals1})-(\ref{Indvals2}). For example, when $a=2$, $b=14$ as in fig.\ \ref{fl214}, we get $I_1 = 1$, $I_2=3729$, $\Omega=3729$.

\subsubsection{D7 tadpole cancelation}

The total D4 charge of our configuration is $2(a+b-2)D$. The total D7 charge of the corresponding IIB configuration is $2(a+b-16)D$. The two are different because $\Gamma_{O7} = 8 \Gamma_{O4}$. In particular, IIB D7 tadpole cancelation requires $a+b=16$, in which case the dual D4-charge equals 28. This discrepancy might seem odd given that we are claiming the two systems are T-dual. But recall that in performing the T-dualities leading from space filling to localized branes, we initially had the spatial $\IR^3$  compactified on a 3-torus. On this 3-torus, there are eight O4-points, so the total D4 charge on the torus is in fact zero, as it should. After decompactification, these additional negative D4 charges due to the seven other O4-planes reside at infinity, and we assume all charge due to D-branes is localized in a finite neighborhood of the O4 point we put at the origin.\footnote{A priori we could have localized the D-brane charges near the other orientifold points as well. The freedom we have to move the center of mass of the brane bound states from one O4 fixed point to another corresponds in the T-dual IIB theory to the freedom of turning on discrete Wilson lines on the D7-branes along the $T^3$ directions. Since this is only possible in the compactified theory, we do not wish to count these possibilities.}

\subsubsection{Microscopic interpretation}

On the IIB side (at large volume and weak string coupling) we have a D7 system carrying certain nontrivial worldvolume fluxes, as described at length in \cite{Collinucci:2008pf}. These systems have rather intricate moduli deformation moduli spaces $\CM$. Recalling the discussion in section \ref{sec:microint}, our result (\ref{indresult}) should give us the Euler characteristic $\chi(\CM)$ of these spaces. Explicitly, for all D7 tadpole canceling configurations with $a<b$, we get the predictions
\vskip3mm
\begin{center}
\begin{tabular}{|r|c|c|c|c|c|c|}
 \hline
 $a$ & 2 & 3 & 4 & 5 & 6 & 7 \\
 \hline
 $\chi(\CM)$ & 3729 & 33540 & 104825 & 223440 & 388905 & 598884 \\
 \hline
\end{tabular}
\end{center}
\vskip4mm

Before verifying this, let us describe these spaces in a little more detail. Starting from the D9-D9$'$ tachyon condensation picture, it was shown in \cite{Collinucci:2008pf} that the moduli spaces $\CM(a,b)$ are given by the space of all supersymmetric tachyon matrices $T$ satisfying the orientifold projection condition, modulo equivalence. The tachyon matrices in the cases at hand are $2 \times 2$ matrices of holomorphic sections $T_{ij}$ of $\CO(n_i) \otimes \sigma^* \CO(n_j) = \CO(n_i+n_j)$ where $n_1=a$, $n_2=b$. A section of $\CO(n)$ is a homogeneous polynomial of the $z_i$ of weight $n$ on the CY given by (\ref{CYeq2}), where $z_0$ has weight 4 and $z_i$ $i \geq 1$ has weight 1. The orientifold projection condition is $T = - \sigma^* T^\top$.
This constrains $T$ to be of the general form
\begin{equation}
T=z_0 \begin{pmatrix}
\rho & \psi \\
\psi & \tau
\end{pmatrix}
+\begin{pmatrix}
0 & \eta\\
-\eta & 0
\end{pmatrix},
\end{equation}
where $\rho, \psi, \tau $ and $\eta$ are homogeneous polynomials in $(z_1,z_2,z_3,z_4)$ of appropriate weight.
The corresponding D7-brane is given by the equation $\det T = 0$, and carries specific $(a,b)$-dependent worldvolume $U(1)$ fluxes as described in detail in section 4.2 of \cite{Collinucci:2008pf}. The equivalence relation to mod out by is
\begin{equation} \label{tacheq}
 T \simeq g \cdot T \cdot \sigma^* g^\top \, ,
\end{equation}
where $g$ is invertible and $g_{ij}$ is a holomorphic section of $\CO(n_i-n_j)$, that is $g_{11}$ and $g_{22}$ are constants, $g_{12}=0$ and $g_{21}$ a polynomial of degree $b-a$. We are assuming here and below that $a < b$. Finally, just as in the description of $\ICP^r$ as a $\IC^*$ quotient of $\IC^{r+1}$ the origin is excluded, here we need to exclude certain values of $T$, namely those corresponding to orbits which do not contain a solution to the D-term constraints. Concretely these are values of $T$ for which the D7 worldvolume $\det T=0$ consists of two components with fluxes carrying nonzero D5 charge on each individual component, such that the two components have different central charge phases and therefore are not mutually supersymmetric.

It is not too hard to see that $\CM$ has the structure of a weighted projective space fibration over a projective space. The fibration projection is $T \to T_{11}$. The base is the space of all nonzero holomorphic sections $T_{11} = z_0 \rho$, modulo $\IC^*$ rescalings. The zero section $T_{11}=0$ is excluded because at this point the D7 splits in two components $\eta = \pm z_0 \psi$ with nonzero flux \cite{Collinucci:2008pf}, breaking supersymmetry. The $\IC^*$ equivalence descends from the $g_{11}$ component of (\ref{tacheq}) (by which we mean we take all components of $g$ to be those of the unit matrix except $g_{11}$). Thus, the base is $\ICP^{\# T_{11} - 1}$,
where $\# T_{11}$ denotes the number of independent holomorphic sections $T_{11}$. The fiber is parametrized by $T_{21}$ and $T_{22}$. The $g_{21}$ component of (\ref{tacheq}) identifies $T_{21}$ and $T_{21} + g_{21} T_{11}$. The inequivalent values of $T_{21}$, $T_{22}$ under this identification thus form a vector space isomorphic to $\IC^{\# T_{21}-\# T_{11}} \oplus \IC^{\# T_{22}}$. Finally the $g_{22}$ component of (\ref{tacheq}) acts as $T_{21} \to g_{22} T_{21}$, $T_{22} \to g_{22}^2 T_{22}$. We conclude that $\CM$ has the following fibration structure:
\begin{equation}
\begin{matrix}
 \mathbb{CP}^{B+C-1}_{\underbrace{2,\cdots,2}_{B},\underbrace{1,\cdots, 1}_{C} }  & \rightarrow&\cal{M}\\
& & \downarrow \\
& & \mathbb{CP}^{A-1}
\end{matrix}
\end{equation}
where
\begin{eqnarray}
 A &=& \# T_{11} = \sm{{2a-4 +3 \choose 3}} = |I(\Gamma_1,\Gamma_0)| \\
 B &=& \# T_{22} = \sm{{2b-4 +3 \choose 3}} = |I(\Gamma_2,\Gamma_0)| \\
 C &=& \# T_{21} - \# g_{21} = \sm{{a+b+3 \choose 3} + {a+b-4+3 \choose 3}  - {b-a +3 \choose 3} - {b-a-4 +3 \choose 3}} = |\langle \Gamma_2,\Gamma_1 + \Gamma_1' \rangle| \, .
\end{eqnarray}
By using the fact that the topological Euler characteristic of a regular fibration equals the product of the Euler characteristics of base and fiber, and the fact that the Euler characteristic of the base equals $A$ while, as shown in appendix \ref{app:euler}, the topological Euler characteristic of the fiber equals $B+C$, we see that the
topological Euler characteristic $\chi(\CM)$ of $\CM$ exactly agrees with the orientihole index (\ref{indresult}). Because the fiber has quotient singularities, it is important to specify what kind of Euler characteristic we are using. In particular, although from (\ref{Omchi}) it was physically expected that \emph{some} notion of Euler characteristic should reproduce the index, the new thing we learn here is that, at least in this case, the index is reproduced by the \emph{topological} Euler characteristic.

The D7 configurations we have studied so far are not yet consistent IIB vacua, since we still have to take into account D3 tadpole cancelation. Introducing mobile D3 branes or worldvolume fluxes to achieve this will increase the number of vacua exponentially. We turn to these in the next part.

\subsection{Tadpole canceling vacua}

Consider again the D7-brane produced by taking $a=2$, $b=14$. This cancels the D7 charge of the O7, but not the D3 charge. To cancel the D3 tadpole, we need to add additional D3 charge, by a combination of worldvolume flux and mobile D3-branes. The O7 has D3-charge $-\frac{152}{3}$ and the $(2,14)$ D7 has D3-charge $-\frac{5680}{3}$, so we need to add $N_{D3}=1944$ units of D3-charge, that is 972 D3-image-D3 pairs. For other values of $a$ we get
\vskip3mm
\begin{center}
\begin{tabular}{|r|c|c|c|c|c|c|c|}
 \hline
 $a$ & 2 & 3 & 4 & 5 & 6 & 7 & 8 \\
 \hline
 $N_{\rm D3}$ & 1944 & 1592 & 1304 & 1080 & 920 & 824 & 792 \\
 \hline
\end{tabular}
\end{center}
\vskip4mm

\subsubsection{Single centered orientiholes}

The total charge of the dual orientihole system will be
\begin{equation}
 \Gamma = - 7 \Gamma_{O4} = 28 \, D - \frac{133}{6} \, D^3 \, .
\end{equation}
Unlike the pure D7 case, this charge has in fact an attractor point:
\begin{equation} \label{totQattr}
 t_*(\Gamma) = \frac{\sqrt{19}}{2} \, i \, D \approx 2.18 \, i \, D \, , \qquad S_{\rm OH}(\Gamma) \approx 1789 \, .
\end{equation}
Here $S_{OH}$ denotes the orientropy, as in (\ref{orientropyval}). If we take this result at face value, we get approximately $e^{1789} \approx 10^{777}$ vacua in this sector. However, we know from the discussion in section \ref{sec:D7flowtrees} that this value of $t_*$ lies outside the regime of validity of the large volume approximation, so genus zero instanton corrections will be important. Moreover the charge is quite small, so higher genus instanton corrections may be important too.  We note however that neglecting instanton corrections amounts to essentially the same approximation as the vacuum counting formulae of \cite{fluxcounting}, in the sense that in both cases, the approximation becomes accurate only in the large D3-charge limit. In the case at hand, we are also counting the contribution to the vacuum degeneracy from mobile D3-branes, so we cannot directly compare to \cite{fluxcounting}, which only took into account flux degrees of freedom. However, in appendix G of \cite{Denef:2007vg}, the techniques and approximations of \cite{fluxcounting} were applied to the problem of counting unorientifolded D4-D0 black hole microstates (including mobile D0 degrees of freedom), indeed resulting in precisely (\ref{OSVintform}) with all instanton corrections dropped, which in saddle point approximation reduces to the black hole entropy in the large volume and large charge approximation. Of course this does not mean that the estimate following from (\ref{totQattr}) is good. Rather, it means that for the purpose of counting D7-brane vacua, the estimate based on the approximations of \cite{fluxcounting} is poor.

For $a>2$, the D0 charge is less than for $a=2$, so the attractor modulus and entropy (\ref{totQattr}) become even smaller, and the situation gets even worse.

Improving the analysis by including instanton corrections will be left for future work.

\subsubsection{Scaling solutions}

\begin{figure}[h]
\begin{center}
\includegraphics[height=3cm]{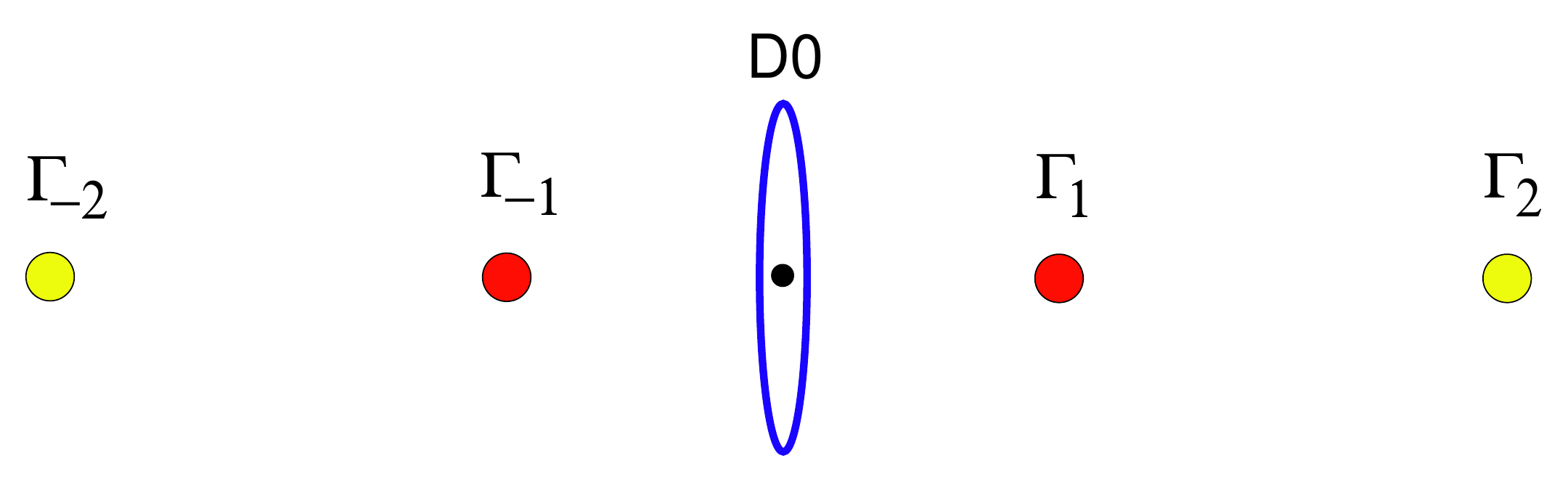}
\end{center}
\caption{Ansatz for scaling solution. The charges $\Gamma_s$ are as in (\ref{ch1})-(\ref{ch2}). The blue ring is equidistant from all other charges $\Gamma_s$ and consists of a large number of D0 particles, not necessarily uniformly distributed. \label{scaling}}
\end{figure}

It is a priori entirely possible that instanton corrections completely destroy the solution (\ref{totQattr}). Since we do not know these corrections, we cannot address this problem directly. In certain circumstances it is nevertheless possible to give a physical argument for existence, namely when the black hole can adiabatically be assembled out of existing ``partons''.  This will be the case when \emph{scaling} solutions exist, similar to those used in \cite{Denef:2007yt} (see also \cite{deconstr}) to ``deconstruct'' D4-D0 black holes. Scaling solutions are solutions to the position constraints (\ref{IntegrabilityConditions}) which persist in the limit in which all position coordinates are uniformly scaled down to zero, or equivalently solutions to (\ref{IntegrabilityConditions}) which persist when $h \equiv 0$. See section 3.8 of \cite{Denef:2007vg} for a review. The idea is to start with a configuration with all partons well separated which we trust exists. Next we let the partons slowly move towards each other. Gravitational backreaction will create a warped throat which grows increasingly long while the position coordinates approach each other. Eventually, the geometry will become indistinguishable from a single centered black hole. In particular, since the process is adiabatic, we expect the single centered black hole solution to exist in the full theory taking into account all corrections.

The simplest possible scaling configuration in the case at hand is shown in fig.\ \ref{scaling}. This immediately generalizes the configurations studied in \cite{Denef:2007yt}. An analysis similar to the one in \cite{Denef:2007yt} shows that a scaling configuration of this kind exists provided the number $N$ of added D0-branes satisfies, for general $(a,b)$:
\begin{equation}
 N \geq N_{\rm sc} = a^3+\frac{2 b^3}{3}+b^2 a-2 a^2-2 b^2+\frac{11 a}{2}+\frac{11 b}{6}-1 \, .
\end{equation}
The critical value is the value of $N$ for which the system of fig.\ \ref{scaling} with the ring collapsed to a point has zero total angular momentum. Indeed scaling solutions necessarily have vanishing angular momentum, since single centered BPS black holes must have zero angular momentum. For the different D7 tadpole canceling cases $(a,16-a)$, this gives
\vskip3mm
\begin{center}
\begin{tabular}{|r|c|c|c|c|c|c|c|}
 \hline
 $a$ & 2 & 3 & 4 & 5 & 6 & 7 & 8 \\
 \hline
 $N_{\rm sc}$ & 1865 & 1682 & 1515 & 1372 & 1261 & 1190 & 1167 \\
 \hline
\end{tabular}
\end{center}
\vskip4mm
Comparing to the table for $N_{\rm D3}$ above, we that for $a=2$, we do get a scaling solution compatible with D3 tadpole cancelation on the IIB side. Thus, by the argument given above, we do expect a single centered orientihole solution dual to tadpole canceling IIB vacua in the full theory.

\subsubsection{Fat multicentered orientiholes}

\begin{figure}[h]
\begin{center}
\includegraphics[height=6cm]{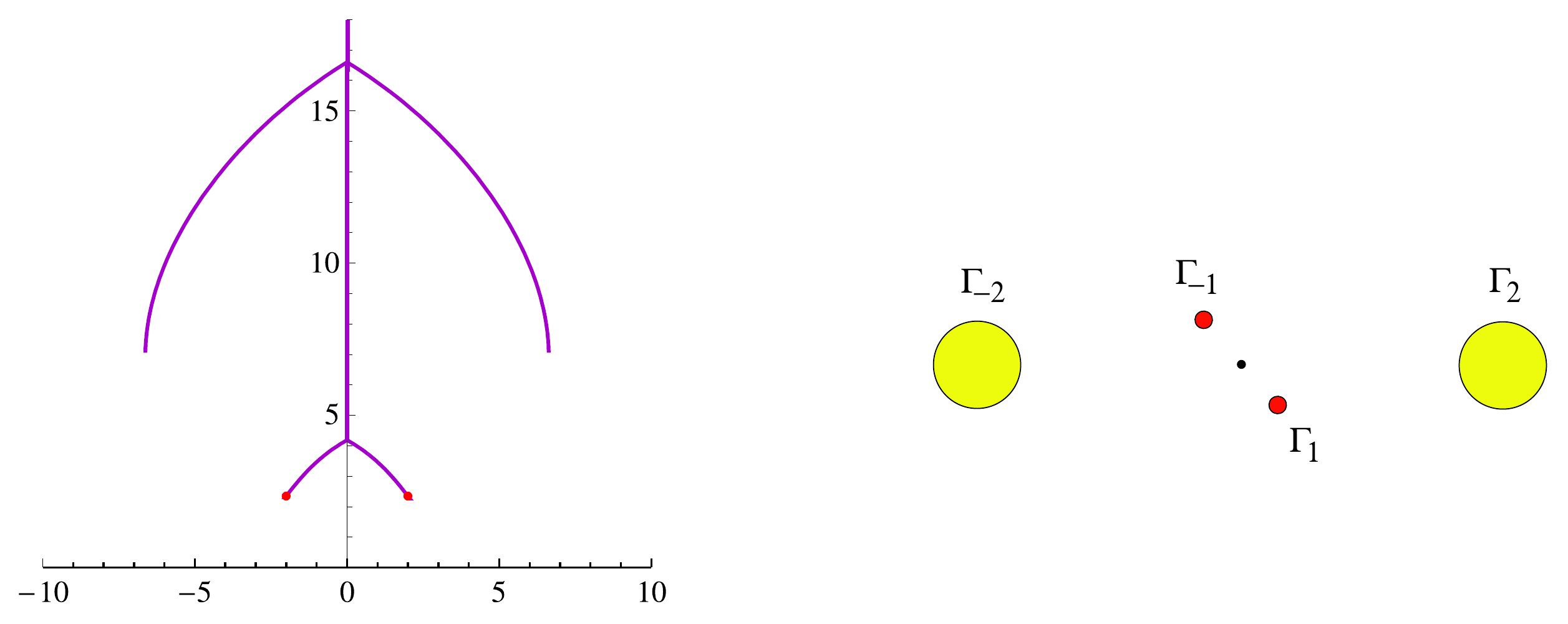}
\end{center}
\caption{Flow tree and configuration for $a=2$, $b=14$, $\beta=53$, $\nu=-256$. \label{fat}}
\end{figure}

We can also consider more general multicentered configurations, for example by replacing (\ref{ch2}) by
\begin{equation}
 \Gamma_2 = e^{b D}\left(1+\sm{\frac{c_2}{24}}-\beta D^2 - \nu D^3\right) \, , \qquad
 \Gamma_{-2} = -e^{-b D}\left(1+\sm{\frac{c_2}{24}}-\beta D^2 + \nu D^3\right) \, .
\end{equation}
This leads to orientihole configurations with nonzero entropy. For example\footnote{The reason we take this particular example is that it is approximately the maximal entropy case within the class under consideration.} for $a=2$, $b=14$, $\beta=53$, $\nu=-256$, we get a IIB tadpole canceling configuration with flow tree shown in fig.\ \ref{fat}. All entropy is carried by $\Gamma_2$, with attractor point and total entropy given by
\begin{equation}
 t_*(\Gamma_2) \approx (6.6 + 7.1 \, i) D \, , \qquad S_{\rm tot} \approx 1540 \, .
\end{equation}
Unlike the single centered case, this lies within the regime of validity of the large volume approximation, as the instanton ratio analogous to (\ref{ratio}) is
\begin{equation} \label{ratio2}
 e^{-2 \pi ({\rm Im}\, t_* - {\rm Im}\, t_0)} \approx 10^{-13} \, .
\end{equation}
Therefore we expect
\begin{equation}
 e^{1540} \approx 10^{668}
\end{equation}
to be a quite reliable lower bound on the number of vacua. Because the charge is not parametrically large, the estimate could presumably still be improved by taking into account subleading corrections in a inverse charge expansion based on a precise version of (\ref{OSVintform}), applied to each of the individual constituent black holes.

\section{Discussion and directions for future research}

We gave a general prescription for constructing orientiholes, discussed their properties, derived a wall crossing formula, conjectured a relation to the topological string, and outlined applications to statistics of the landscape of D7 vacua.

We left several problems untouched:
\begin{itemize}

\item We did not discuss the lift to M-theory. One question one could ask is whether the singular solutions sourced by O4-planes get resolved in M-theory, in the way O6$^-$-planes get resolved by the Atiyah-Hitchin manifold.

\item We analyzed orientiholes in $\IR^3$, but for IIB tadpole canceling configurations, it should be possible to construct solutions on a compactified space $T^3$. When the distance of the charges to an orientifold plane is much less than the size of the torus, the solutions will be approximately the $\IR^3$ ones. But when approaching a wall of marginal stability, some charges will stray far from home, and the solutions will be qualitatively different. In fact, ``decay'' at marginal stability appears no longer possible, since the charges can't leak out to infinity in a compact space. On the other hand, from the classical microscopic picture, one still expects BPS states to disappear when crossing the wall. This is puzzling.

\item The effect of instanton corrections on our vacuum counting estimates needs to be investigated, and a more systematic exploration of the space of IIB tadpole canceling configurations would be desirable.

\item It is plausible that qualitatively different kinds of black hole configurations correspond to qualitatively different sectors of the IIB landscape of D7 vacua. For example, multicentered bound states with nonzero angular momentum appear to correspond to sectors of the D7 landscape with unfixed moduli. This is because microscopically, rotational spin is identified with Lefshetz spin, which is zero on zero dimensional components of the space of all supersymmetric D7 configurations. Furthermore, nonabelian sectors should correspond to configurations with identical particles. Mapping out the relation between these sectors and using this to obtain relative abundances of vacua in different sectors is an important problem.

\item Many of the D7 configurations we studied decay already at relatively large Calabi-Yau radii, significantly larger than those typically obtained in the moduli stabilization scenario of \cite{Kachru:2003aw}. For example, from fig.\ \ref{fl214} we see that the pure D7 decays at a volume $\frac{v^3}{6} \approx 4600$ in string units.
    This shows in particular that classical geometry breaks down dramatically already at these relatively large scales, casting some doubt on the reliability of much of the work on string phenomenology based on classical geometry.

\item We derived a wall crossing formula (\ref{wcf}) only for primitive charges $\Gamma_0$ and $\Gamma_1$. The next step would be to derive such a formula for halos with several copies of $\Gamma_1$. The complicating factor here is that although the copies of $\Gamma_1$ don't interact among themselves, they do interact with each others images. More ambitiously, one could ask if there is an orientihole generalizion of \cite{KS,Gaiotto:2008cd,Nis4}.

\item We did not provide a proof for (a version of) the OSV-like conjecture (\ref{DEPintform}). Something along the lines of \cite{Denef:2007vg,Gaiotto:2006ns,deBoer:2006vg} might work.

\item One could ask if exact results for orientihole partition functions, analogous to \cite{Gaiotto:2006wm,Collinucci:2008ht,Dijkgraaf:1996it}, can be obtained, and what their modularity properties are.

\item We have ignored bulk fluxes on the IIB side, such as RR 3-form fluxes. The latter are sourced by D5 domain walls wrapping 3-cycles. These T-dualize to D4 strings wrapping 3-cycles. Thus one would be led to consider bound states of such strings and black holes. This is somewhat reminiscent of \cite{Aganagic:2005dh}.

\item It is natural to ask if the relation between black hole microstates and D-brane vacua can be exploited in the absence of supersymmetry.

\end{itemize}

\section*{Acknowledgements}

We would like to thank Greg Moore for useful discussions. This work is supported in part by DOE grant DE-FG02-91ER40654 and a DOE OJI award.

%\newpage

\appendix

\section{Explicit expressions in prolate spheroidal coordinates} \label{app:explicit}

We introduce the prolate spheroidal coordinates  $(\rho_s, \psi_s,\phi_s)$ with foci at   $\vec{x}_s$ and $-\vec{x}_{s}$. The angle  $\phi_s\in[0,2\pi]$ is still the azimuthal angle around the axis $z_s$.
The angle $\psi_s\in[0,\pi]$ and the coordinates
  $\rho_s>0$   are defined by
\begin{equation}
r_s+r_{-s}=2a_s \cosh \rho_s,\\ \quad
r_s-r_{-s}=2a_s \cos \psi_s.
\end{equation}
where $r_s$ and $r_{-s}$ are the distances from $\vec{x}$ to $\vec{x}_s$ and $-\vec{x}_{s}$, respectively.
The surfaces of constant $\rho_s$ form  {\em  prolate spheroids}  obtained by rotating an ellipse with foci $\vec{x}_s$ and $-\vec{x}_{s}$  around the $z_s$-axis. A surface of constant $\psi_s$ is a hyperboloid of revolution.   We  have the following relations :
\begin{equation}
r_s r_{-s} =\frac{(r_s+r_{-s})^2-(r_s-r_{-s})^2}{4}=a_s^2(\cosh^2 \rho_s -\cos^2 \psi_s)=a_s^2(\sinh^2 \rho_s +\sin^2 \psi_s).
\end{equation}

The link with a Cartesian system centered at the origin with the $z$-axis identified with the $z_s$-axis  is given by
\begin{eqnarray}
x &=& a_s \sinh \rho_s \sin \psi_s \cos \phi_s \\
y &=& a_s \sinh \rho_s \sin \psi_s \sin \phi_s \\
z &=& a_s \cosh \rho_s \cos \psi_s \phantom{\sin \phi_s}
\end{eqnarray}
Therefore, the parity operator acts as follows:
\begin{equation}
\label{ParityProlateSpheroidal}
\mathscr{P}:\quad \psi_s\mapsto \pi-\psi_s,\quad   \phi_s\mapsto \phi_s+\pi, \quad  \rho_s \mapsto \rho_s.\end{equation}

The off-diagonal part of the metric can be computed explicitly by solving the equation  $d\omega=\star\langle dH, H\rangle$. We present below the two-centered solution for a hole and its orientifold image at $\vec{x}_s$ and $-\vec{x}_s$. The geometry is shown in Figure \ref{Figure.2centers}. The most general solution for $\omega$ is simply a superposition of such expressions for every pair of centers in the configuration.

Using the integrability condition and the antisymmetry of the symplectic metric, we have
\begin{align}
\langle dH, H\rangle=\frac{1}{2}
  \langle  \Gamma_s , \Gamma_{-s} \rangle \left(r_{-s}^{-1} dr_s^{-1} -r_s^{-1} dr_{-s}^{-1}+r^{-1} dr^{-1}_{-s}- r^{-1} dr^{-1}_s \right),
\end{align}
where $r_s$ and $r_{-s}$ are the distances from $\vec{x}$ to the hole and image hole, respectively.
The solution is given by
\begin{align}
\omega=\frac{\langle \Gamma_s, \Gamma_{-s} \rangle}{r}\left({
\frac{a_s^2-r^2  }{\sqrt{r_s^4+r_{-s}^4-2 r_s^2 r_{-s}^2 \cos 2\theta)}}+1-\cos\theta_s +\cos  \theta_{-s}   }
\right)  d\phi_s.
\end{align}

Expressing the angles in terms of the distances $r$, $a_s$,  $r_s$ and $r_{-s}$, we get
\begin{align}
\omega=\frac{\langle \Gamma_s, \Gamma_{-s} \rangle}{r}\frac{(4 a_s^2-(r_s-r_{-s})^2)(
2a_s-(r_s+r_{-s})) }{4 a_s r_s  r_{-s}} d\phi_s.
\end{align}
Writing it in prolate spheroidal coordinates, we find that
\begin{equation}
\omega
=-4 \langle \Gamma_s, \Gamma_{-s} \rangle    \frac{\sinh^2 \frac{\rho_s}{2} \sin^2 \psi_s}{\sinh^2 \rho_s +\sin^2 \psi_s} d\phi_s.
\end{equation}
Checking with (\ref{ParityProlateSpheroidal}), we see that $\omega$ is even under parity, as required by the orientifold projection.

Similarly, the gauge fields can be written in prolate spheroidal coordinates as
 \begin{align}
  {\cal A}^{ \Lambda+}_{\mathscr{D}} &=  P^{ \Lambda +}_0(c_0-\cos \theta) d\phi+\sum_s  {P^{ \Lambda+}_s}\Big(c^+_s -\frac{2 \sinh^2 \rho_s    \cos \psi_s}{\sinh^2 \rho_{s} +\sin^2 \psi_{s}}\Big)d\phi_s, \\
  {\cal A}^{ \Lambda-}_{\mathscr{D}} &=\sum_s {P^{ \Lambda-}_s}\Big(c_s^--2 \frac{  \sin^2 \psi_s \cosh \rho_s}{\sinh^2 \rho_{s} +\sin^2 \psi_{s}} \Big)  d\phi_s.
 \end{align}
  The parity transformation  rules for the constants are :
\begin{equation}
\mathscr{P}:
 c_s^\pm \mapsto \mp  c_s^\pm, \quad c_0  \mapsto   -c_0 .
\end{equation}
We see that $ \mathscr{P}
  {\cal A}^{\Lambda \pm}_{\mathscr{D} }=
  \mp {\cal A}^{\Lambda\pm}_{\mathscr{D}}
$. That is,
\begin{align}
 \mathscr{P}:\qquad
  ({\cal A}^{A\pm}_{\mathscr{D} },
    {\cal A}^{0}_{\mathscr{D} })
  \quad   \rightarrow\quad (-)^\epsilon ( \mp {\cal A}^{A\pm}_{\mathscr{D}},  -{\cal A}^0_{\mathscr{D} }),
\end{align}
as it should.

\section{Gauge bundle derivation of Dirac quantization and torsion charges} \label{app:gaugebundle}

Instead of working with unique vector potentials and Dirac strings, one can also cover space with patches carrying different, nonsingular vector potentials, related by gauge transformations. The gauge transformations define the transition functions of a vector bundle.

\begin{figure}[!tbh]
\begin{center}
\begin{picture}(40,80)
\thicklines
\put(0,0){\circle{2}}
\put(40,0){\circle{2}}
\put(80,0){\circle{2}}
\put(100,60){\circle{2}}
\put(-20,0){\line(1,0){120}}
\put(0,0){\line(5,3){100}}
\put(80,0){\line(1,3){20}}
\put(10,3){\oval(2,6)[r]}
\put(81.5,3.5){\oval(9,6)[r]}
\put(44.5,3){\oval(6,6)[r]}
\put(40,0){\line(1,1){60}}
\put(-10,-10){\footnotesize \text{$-\vec x_s$}}
\put(38,-12){\footnotesize \text{$\vec{0}$}}
\put(60,-10){\footnotesize \text{$a_s$}}
\put(80,-10){\footnotesize \text{$\vec x_{s}$}}
\put(100,65){$\vec x$}
\put(17,3){\footnotesize \text{$\theta_{-s}$}}
\put(50,3){\footnotesize \text{$\theta$}}
\put(88,5){\footnotesize \text{$\theta_{s}$}}
\put(23,25){\footnotesize \text{$\vec r_{-s}$}}
\put(68,20){\footnotesize \text{$\vec r$}}
\put(92,25){\footnotesize \text{$\vec r_{s}$}}
\end{picture}
\caption{  Two black holes located at  $\vec x_s$ and $-\vec x_{s}$. The middle point is the origin. \label{Figure.2centers} }
\end{center}
\end{figure}
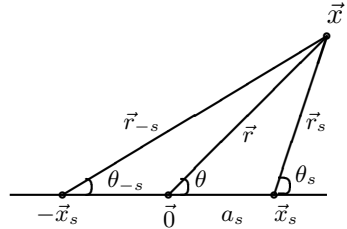

The topologically nontrivial information is completely contained in the Dirac part ${\cal A}^\Lambda_\mathscr{D}$ of the gauge potentials.  Using linearity,  ${\cal A}^\Lambda_\mathscr{D}$ can be expressed   as a superposition of solutions for each individual charge.
The BPS equation of motion  $d {\cal  A}^\Lambda_{\mathscr{D},s}=-P^\Lambda_s (\star d\tau_s)$  admits as a solution a  magnetic monopole  centered at  $\vec x_s$ and carrying a magnetic charge $P^\Lambda_s$:
$
{\cal A}^\Lambda_{\mathscr{D},s}=P^\Lambda_s  (c_s-\cos \theta_{s}) d\phi_s$.
   The angle $\theta_s$ is shown in Figure \ref{Figure.2centers}. The $z_s$ axis is defined as the line connecting the $s$th center and its image.
The   $c_s$   is a  constant of integration  carefully chosen in different patches  to cancel the singularity of the one-form $d\phi_s$ along the $z_s$-axis. For $\epsilon=0$, we introduce the notation $\Lambda+=A+$, $\Lambda- \in \{0,A-\}$. For $\epsilon=1$ on the other hand we denote $\Lambda+ \in \{0,A-\}$, $\Lambda+=A+$. Then $\mathscr{P}{\cal A}^{\Lambda \pm}_{\mathscr{D} }=\mp {\cal A}_{\mathscr{D} }^{\Lambda \pm}$, and if the $s$th-hole has  magnetic charge  $P^{\Lambda+}_s$ ($P^{\Lambda-}_s$) its image has magnetic charge $+ P^{\Lambda+}_s$ ($- P^{\Lambda-}_s$). Grouping a  hole and its image-hole, we obtain:
 \begin{align}
 \label{EvenChargeGaugeField}
  {\cal A}^{\Lambda+}_{\mathscr{D}} &=P^{\Lambda+}_0(c_{0}-\cos \theta) d\phi+\sum_s P^{\Lambda+}_s  \big(c_{s}^+-\cos \theta_s -\cos \theta_{-s}\big)d\phi_s, \\
  \label{OddChargeGaugeField}
  {\cal A}^{\Lambda-}_{\mathscr{D}} &=\sum_s  P^{\Lambda-}_s\big(c_s^--\cos \theta_s +\cos \theta_{-s}\big)d\phi_s, \end{align}
  where the constants $c_0$ and $c_s^\pm$ are given by
$ c_0={\rm Sign}(z_s)$ and $c_s^\pm=
{\rm Sign} (z_s-a_s)\pm {\rm Sign}(z_s+a_s)
$, that is
 \begin{center}
\begin{picture}(350,60)
\thicklines
\put(10,50){\line(1,0){340}}
\put(350,50){\line(-1,1){10}}
\put(350,50){\line(-1,-1){10}}
\multiput(90,50)(80,0){3}{\circle*{5}}
\put(80,55){$-a_s$}
\put(168,55){$0$}
\put(240,55){$+a_s$}
\put(350,60){\text{$z_s$-axis}}
\put(15,0){\line(0,1){60}}
\put(0,20){$\begin{matrix} c^-_s\\ c^0_s \\ c^+_s\end{matrix}$}
\put(40,20){$\begin{matrix} \phantom{-}0\\ -1\\ -2\end{matrix}$}
\put(110,20){$\begin{matrix} -2\\ -1\\ \phantom{-}0\end{matrix}$}
\put(200,20){$\begin{matrix} -2\\ +1\\ \phantom{-}0\end{matrix}$}
\put(290,20){$\begin{matrix} \phantom{-}0\\ +1\\ +2\end{matrix}$}
\end{picture}
\end{center}

\subsection{Transition functions and Dirac quantization}

Let us now consider the transition functions in more detail and rederive the quantization conditions (\ref{DiracQuant}) in this framework.
We focus on the charge at the origin and the $s$-th charge-image-charge pair, and on some particular component of the gauge field. A natural choice of patches for ${\cal A}^{\Lambda+}$  divides space in four overlapping regions $U_-,U_{0-},U_{0+},U_+$ on which the different vector potentials given above are well defined. Away from the $z=z_s$-axis, $U_-$ contains $z<-a_s$, $U_{0-}$ contains $-a_s<z<0$, $U_{0+}$ contains $0<z<a_s$ and $U_+$ contains $a_s<a$.  For ${\cal A}^{\Lambda-}$, we need only three patches $U_-,U_{0},U_+$ where $U_\pm$ are defined as before and $U_0$ is the union of $U_{0+}$ and $U_{0-}$, that is  $U_{0}$ contains $-a_s<z<a_s$.
We can take $\mathscr{P} U_{-} = U_+$ and $\mathscr{P} U_{0-} = U_{0+}$. The corresponding transition functions are all of the form
\begin{equation}\label{gt}
{\cal A}\mapsto {\cal A}-ig^{-1}dg,\quad
\text{where} \quad g=\exp\big[i m(\phi+\phi_0)\big], \quad m\in \mathbb{Z}, \phi_0\in \mathbb{R},
\end{equation}
where $\phi$ is the azimuthal angle. The constant $m$ is required to be an integer $(m\in \mathbb{Z})$ in order to ensure that the transition function is single-valued. However, for a transition function in the neighborhood of the origin (i.e.\ the one between $U_{0-}$ and $U_{0+}$), the gauging of parity requires that $g$ is single valued under $\phi\mapsto \phi+\pi$. It follows that we should have $m\in 2\mathbb{Z}$ in this case.
 The constant  $\phi_0$ would be irrelevant in the unorientifolded case since it cancels out in  equation \eqref{gt}, but it is convenient to keep it in the orientifolded case. The transition functions should be compatible with the symmetry of the gauge field under parity.  A transition function  $g_{{}_{U,U'}}$ from a region $U$ to a region  $U'$, should be related to the transition function  $g_{{}_{\mathscr{P}U,\mathscr{P}U'}}$   defined between the image  region $\mathscr{P} U$ and $\mathscr{P}U'$:
\begin{equation}\label{gpc}
\mathscr{P}{\cal A}^{\Lambda \pm}_{\mathscr{D} }=\mp {\cal A}_{\mathscr{D} }^{\Lambda \pm} \quad \Leftrightarrow \quad
\mathscr{P}g_{{}_{U,U'}}=
\big(g_{{}_{\mathscr{P} U, \mathscr{P}U'}}\big)^{\mp 1}.
\end{equation}
%We see that we can always satisfy such conditions when $U\cap U'$ is away from the origin (i.e.\ for $(U_{-},U_{0-})$ and $(U_{+},U_{0+})$ transition functions) by taking
%$g_{{}_{U,U'}}=\exp(i m\phi)$ and $
%g_{{}_{\mathscr{P} U,\mathscr{P}U'}}=\exp\big[\pm i m(\phi+\pi )\big]$.
%In the neighborhood of the origin, it is always possible to find an open set $U$ intersecting its image ${\mathscr{P}}U$. If there is a transition function at the origin and an odd gauge field,  as is the case for ${\cal A}^\Lambda_{\mathscr{D},0}$,  we must have   $m\in 2\mathbb{Z}$.
For the $U(1)$ labeled by $\Lambda\pm$, the following transition functions are consistent with all requirements:
 \begin{center}
\begin{picture}(350,80)
\thicklines
\color{blue}
\put(90,50){\circle*{5}}
\put(250,50){\circle*{5}}
\color{black}
\put(10,50){\line(1,0){340}}
\put(350,50){\line(-1,1){10}}
\put(350,50){\line(-1,-1){10}}
\put(40,60){$U_{-}$}
\put(168,60){$U_{0}$}
\put(300,60){$U_{+}$}
\put(240,55){$+a_s$}
\put(80,55){$-a_s$}
\put(45,25){$\begin{matrix}g_{{}_{U_-U_0}}\\= \\\exp[- i P^{\Lambda-}_s \phi]\end{matrix}$}
\put(200,25){$\begin{matrix}g_{{}_{U_0U_+}}\\=\\\exp\big[ i e P^{\Lambda-}_s (\phi+\pi)\big]\end{matrix}$}
\put(350,60){\text{$z_s$-axis}}
\end{picture}
\end{center}
 \begin{center}
\begin{picture}(350,80)
\thicklines
\put(0,50){\line(1,0){350}}
\put(350,50){\line(-1,1){10}}
\put(350,50){\line(-1,-1){10}}
\color{red}
\put(270,50){\circle*{5}}
\put(50,50){\circle*{5}}
\color{yellow}
\put(160,50){\circle*{5}}
\color{black}
\put(0,60){$U_{-}$}
\put(110,60){$U_{0-}$}
\put(218,60){$U_{0+}$}
\put(310,60){$U_{+}$}
\put(40,55){$-a_s$}
\put(260,55){$+a_s$}
\put(158,55){$0$}
\put(15,25){$\begin{matrix} g_{{}_{U_-U_{0-}}}\\=\\\exp(i P^{\Lambda+}_s \phi)\end{matrix}$}
\put(130,25){$\begin{matrix}
g_{{}_{U_{0-}U_{0+}}}\\=\\
\exp(i P^{\Lambda+}_0 \phi)\end{matrix}$}
\put(220,25){$\begin{matrix}
g_{{}_{U_{0+}U_{+}}}\\=\\\exp\big[i P^{\Lambda+}_s (\phi+\pi) \big]\end{matrix}$}
\put(350,60){\text{$z_s$-axis}}
\end{picture}
\end{center}
where
\begin{equation}
  P^{\Lambda }_0 \in 2 \mathbb{Z}, \qquad P^{\Lambda \pm}_s \in \mathbb{Z}.
\end{equation}
This reproduces (\ref{DiracQuant}).

\subsection{$\mathbb{Z}_2$-torsion charge}

It is straightforward to compute the Wilson lines determining the $\IZ_2$ torsion charges defined in section \ref{sec:torsion}. A unit probe electric charge dual to $P^{\Lambda-}$ magnetic charge, circling the path $\alpha$, picks up a phase:
\begin{equation}
 e^{i \Phi(\alpha)} = e^{\frac{i}{2} \oint_{\alpha} {\cal A}^{\Lambda-}_{\mathscr{D},s} }
=e^{\sum_{s>0} \frac{i}{2} \cdot (-2P^{\Lambda-}_s) \cdot \pi} =
e^{i \pi \sum_{s>0} P^{\Lambda-}_s} \, ,
\end{equation}
reproducing (\ref{torsionformula}).

\section{Topological Euler characteristic of  $\mathbb{CP}^{m+n-1}_{1,\cdots,1,r,\cdots, r}$ } \label{app:euler}

 The topological Euler characteristic of $\mathbb{CP}^{m+n-1}_{1,\cdots,1,r,\cdots, r}$ can be computed as follows. First we define a continuous surjective map from the  projective space $\mathbb{CP}^{m+n-1}$ to the weighted projective space $\mathbb{CP}^{m+n-1}_{1,\cdots,1,r,\cdots, r}$. Let us denote the projective coordinates of
  $\mathbb{CP}^{m+n-1}$ by $[x_i,y_j]$; then the  mapping is defined as
  \begin{equation}
\mathbb{CP}^{m+n-1}\longrightarrow
\mathbb{CP}^{m+n-1}_{1,\cdots,1,r,\cdots, r}:
  [x_i, y_i]\mapsto [x_i, y_j^r].
  \end{equation}
  The mapping is clearly surjective but is not in general one-one since the point  $[x_i,y_j]$
 and $[x_i, e^{\frac{2\pi }{r} i } y_j]$  which are different in $\mathbb{CP}^{m+n-1}$ (for $\vec x\neq 0$ and $\vec y\neq \vec 0$)  are mapped to the same point of   $\mathbb{CP}^{m+n-1}_{1,\cdots,1,r,\cdots, r}$.
  So $\mathbb{CP}^{m+n-1}$ is a multi-covering of $\mathbb{CP}^{m+n-1}_{1,\cdots,1,r,\cdots, r}$. However, the number of branches is not constant: the map is  $r:1$ for
  $\vec x\neq \vec 0$ and $\vec y\neq \vec 0$, but the map is one-one when $\vec x=\vec 0$ or $\vec y=\vec 0$. Keeping track of the topology we get:

 \begin{center}
 \begin{tabular}{|  c |  c | c |}
 \hline
 Sector of  $\mathbb{CP}^{m+n-1}$ & Number of branches & Topology \\
 \hline
 $\vec x \neq \vec 0$ and $\vec y\neq \vec 0$& $r$  & $\mathbb{CP}^{m+n-1}-\mathbb{CP}^{m-1}-\mathbb{CP}^{n-1}$\\
 \hline
 $\vec x \neq \vec 0$ and $\vec y=\vec 0$& $1$  &
 $\mathbb{CP}^{m-1}$
 \\
 \hline
 $\vec x = \vec 0$ and $\vec y\neq \vec 0$& $1$  & $\mathbb{CP}^{n-1}$\\
 \hline
 \end{tabular}
\end{center}

We can then compute straightforwardly  the  topological Euler characteristic by adding the pieces together with the appropriate weights:
 \begin{align}
 \chi_{top}(\mathbb{CP}^{m+n-1}_{1,\cdots, 1, r,\cdots, r }) &=
 \frac{1}{r}\chi(\mathbb{CP}^{m+n-1}-\mathbb{CP}^{m-1}-\mathbb{CP}^{n-1})+
 \chi(\mathbb{CP}^{m-1})+\chi(\mathbb{CP}^{n-1})\nonumber\\
 &=
 \frac{1}{r}\Big[\chi(\mathbb{CP}^{m+n-1})-\chi(\mathbb{CP}^{m-1})-\chi(\mathbb{CP}^{n-1})\Big]+
 \chi(\mathbb{CP}^{m-1})+\chi(\mathbb{CP}^{n-1})\nonumber \\
 &=m+n.
  \end{align}


\begin{thebibliography}{1}

%\cite{Kachru:2003aw}
\bibitem{Kachru:2003aw}
  S.~Kachru, R.~Kallosh, A.~Linde and S.~P.~Trivedi,
  ``De Sitter vacua in string theory,''
  Phys.\ Rev.\  D {\bf 68}, 046005 (2003)
  [arXiv:hep-th/0301240].
  %%CITATION = PHRVA,D68,046005;%%

%\cite{Douglas:2006es}
\bibitem{Douglas:2006es}
  M.~R.~Douglas and S.~Kachru,
  ``Flux compactification,''
  Rev.\ Mod.\ Phys.\  {\bf 79}, 733 (2007)
  [arXiv:hep-th/0610102].\\
  %%CITATION = RMPHA,79,733;%%

%\cite{Beasley:2008dc}
\bibitem{Beasley:2008dc}
  C.~Beasley, J.~J.~Heckman and C.~Vafa,
  ``GUTs and Exceptional Branes in F-theory - I,''
  arXiv:0802.3391 [hep-th].
  %%CITATION = ARXIV:0802.3391;%%

%\cite{Beasley:2008kw}
\bibitem{Beasley:2008kw}
  C.~Beasley, J.~J.~Heckman and C.~Vafa,
  ``GUTs and Exceptional Branes in F-theory - II: Experimental Predictions,''
  arXiv:0806.0102 [hep-th].
  %%CITATION = ARXIV:0806.0102;%%

%\cite{Heckman:2008rb}
\bibitem{Heckman:2008rb}
  J.~J.~Heckman and C.~Vafa,
  ``From F-theory GUTs to the LHC,''
  arXiv:0809.3452 [hep-ph].
  %%CITATION = ARXIV:0809.3452;%%

%%\cite{Blumenhagen:2008zz}
\bibitem{Blumenhagen:2008zz}
  R.~Blumenhagen, V.~Braun, T.~W.~Grimm and T.~Weigand,
  ``GUTs in Type IIB Orientifold Compactifications,''
  arXiv:0811.2936 [hep-th].
  %%CITATION = ARXIV:0811.2936;%%

%\cite{Blumenhagen:2006ci}
\bibitem{Blumenhagen:2006ci}
  R.~Blumenhagen, B.~Kors, D.~Lust and S.~Stieberger,
   ``Four-dimensional String Compactifications with D-Branes, Orientifolds and
  Fluxes,''
  Phys.\ Rept.\  {\bf 445}, 1 (2007)
  [arXiv:hep-th/0610327].
  %%CITATION = PRPLC,445,1;%%


%\cite{Baumann:2007ah}
\bibitem{Baumann:2007ah}
  D.~Baumann, A.~Dymarsky, I.~R.~Klebanov and L.~McAllister,
  ``Towards an Explicit Model of D-brane Inflation,''
  JCAP {\bf 0801}, 024 (2008)
  [arXiv:0706.0360 [hep-th]].
  %%CITATION = JCAPA,0801,024;%%

%\cite{Chen:2009nk}
\bibitem{Chen:2009nk}
  H.~Y.~Chen, L.~Y.~Hung and G.~Shiu,
  ``Inflation on an Open Racetrack,''
  arXiv:0901.0267 [hep-th].
  %%CITATION = ARXIV:0901.0267;%%



 %\cite{Collinucci:2008pf}
\bibitem{Collinucci:2008pf}
  A.~Collinucci, F.~Denef and M.~Esole,
  ``D-brane Deconstructions in IIB Orientifolds,''
  arXiv:0805.1573 [hep-th].
  %%CITATION = ARXIV:0805.1573;%%

%\cite{Denef:2008wq}
\bibitem{Denef:2008wq}
  F.~Denef,
  ``Les Houches Lectures on Constructing String Vacua,''
  arXiv:0803.1194 [hep-th].
  %%CITATION = ARXIV:0803.1194;%%

%\cite{Douglas:2003um}
\bibitem{Douglas:2003um}
  M.~R.~Douglas,
  ``The statistics of string/M theory vacua,''
  JHEP {\bf 0305}, 046 (2003)
  [arXiv:hep-th/0303194].
  %%CITATION = JHEPA,0305,046;%%

%\cite{Bousso:2000xa}
\bibitem{Bousso:2000xa}
  R.~Bousso and J.~Polchinski,
  ``Quantization of four-form fluxes and dynamical neutralization of the
  cosmological constant,''
  JHEP {\bf 0006}, 006 (2000)
  [arXiv:hep-th/0004134].
  %%CITATION = JHEPA,0006,006;%%


\bibitem{fluxcounting}

  S.~Ashok and M.~R.~Douglas,
  ``Counting flux vacua,''
  JHEP {\bf 0401}, 060 (2004)
  [arXiv:hep-th/0307049].
  %%CITATION = JHEPA,0401,060;%%

  F.~Denef and M.~R.~Douglas,
  ``Distributions of flux vacua,''
  JHEP {\bf 0405}, 072 (2004)
  [arXiv:hep-th/0404116].
  %%CITATION = JHEPA,0405,072;%%

  F.~Denef and M.~R.~Douglas,
  ``Distributions of nonsupersymmetric flux vacua,''
  JHEP {\bf 0503}, 061 (2005)
  [arXiv:hep-th/0411183].
  %%CITATION = JHEPA,0503,061;%%

  B.~S.~Acharya, F.~Denef and R.~Valandro,
  ``Statistics of M theory vacua,''
  JHEP {\bf 0506}, 056 (2005)
  [arXiv:hep-th/0502060].
  %%CITATION = JHEPA,0506,056;%%

   M.~R.~Douglas, B.~Shiffman and S.~Zelditch,
  ``Critical points and supersymmetric vacua. III: String/M models,''
  Commun.\ Math.\ Phys.\  {\bf 265}, 617 (2006)
  [arXiv:math-ph/0506015].
  %%CITATION = CMPHA,265,617;%%

  F.~Denef,
  ``Les Houches Lectures on Constructing String Vacua,''
  arXiv:0803.1194 [hep-th].
  %%CITATION = ARXIV:0803.1194;%%


\bibitem{opencount}

  R.~Blumenhagen, F.~Gmeiner, G.~Honecker, D.~Lust and T.~Weigand,
  ``The statistics of supersymmetric D-brane models,''
  Nucl.\ Phys.\  B {\bf 713}, 83 (2005)
  [arXiv:hep-th/0411173].
  %%CITATION = NUPHA,B713,83;%%

  F.~Gmeiner, R.~Blumenhagen, G.~Honecker, D.~Lust and T.~Weigand,
  ``One in a billion: MSSM-like D-brane statistics,''
  JHEP {\bf 0601}, 004 (2006)
  [arXiv:hep-th/0510170].
  %%CITATION = JHEPA,0601,004;%%

  F.~Gmeiner and M.~Stein,
  ``Statistics of SU(5) D-brane models on a type II orientifold,''
  Phys.\ Rev.\  D {\bf 73}, 126008 (2006)
  [arXiv:hep-th/0603019].
  %%CITATION = PHRVA,D73,126008;%%

  M.~R.~Douglas and W.~Taylor,
  ``The landscape of intersecting brane models,''
  JHEP {\bf 0701}, 031 (2007)
  [arXiv:hep-th/0606109].
  %%CITATION = JHEPA,0701,031;%%



%\cite{Denef:2000nb}
\bibitem{Denef:2000nb}
  F.~Denef,
  ``Supergravity flows and D-brane stability,''
  JHEP {\bf 0008}, 050 (2000)
  [arXiv:hep-th/0005049].
  %%CITATION = JHEPA,0008,050;%%

%\cite{Behrndt:1997ny}
\bibitem{Behrndt:1997ny}
  K.~Behrndt, D.~Lust and W.~A.~Sabra,
  ``Stationary solutions of N = 2 supergravity,''
  Nucl.\ Phys.\ B {\bf 510}, 264 (1998)
  [arXiv:hep-th/9705169].
  %%CITATION = HEP-TH 9705169;%%

%\cite{LopesCardoso:2000qm}
\bibitem{LopesCardoso:2000qm}
  G.~Lopes Cardoso, B.~de Wit, J.~Kappeli and T.~Mohaupt,
``Stationary BPS solutions in N = 2 supergravity with R**2
interactions,''
  JHEP {\bf 0012}, 019 (2000)
  [arXiv:hep-th/0009234].
  %%CITATION = HEP-TH 0009234;%%

  %\cite{Denef:2002ru}
\bibitem{Denef:2002ru}
  F.~Denef,
  ``Quantum quivers and Hall/hole halos,''
  JHEP {\bf 0210}, 023 (2002)
  [arXiv:hep-th/0206072].
  %%CITATION = JHEPA,0210,023;%%

\bibitem{Denef:2007vg}
F.~Denef and G.~W.~Moore,
``Split states, entropy enigmas, holes and halos,''
arXiv:hep-th/0702146.
%%CITATION = HEP-TH/0702146;%%


%\cite{Ooguri:2004zv}
\bibitem{Ooguri:2004zv}
  H.~Ooguri, A.~Strominger and C.~Vafa,
  ``Black hole attractors and the topological string,''
  Phys.\ Rev.\  D {\bf 70}, 106007 (2004)
  [arXiv:hep-th/0405146].
  %%CITATION = PHRVA,D70,106007;%%


  %\cite{Bates:2003vx}
\bibitem{Bates:2003vx}
  B.~Bates and F.~Denef,
  ``Exact solutions for supersymmetric stationary black hole composites,''
  arXiv:hep-th/0304094.
  %%CITATION = HEP-TH/0304094;%%


\bibitem{Acharya:2002ag}
  B.~S.~Acharya, M.~Aganagic, K.~Hori and C.~Vafa,
  ``Orientifolds, mirror symmetry and superpotentials,''
  arXiv:hep-th/0202208.
  %%CITATION = HEP-TH/0202208;%%


%\cite{Brunner:2003zm}
\bibitem{Brunner:2003zm}
  I.~Brunner and K.~Hori,
  ``Orientifolds and mirror symmetry,''
  JHEP {\bf 0411}, 005 (2004)
  [arXiv:hep-th/0303135].
  %%CITATION = JHEPA,0411,005;%%




\bibitem{Polchinski:1996fm}
  J.~Polchinski, S.~Chaudhuri and C.~V.~Johnson,
  ``Notes on D-Branes,''
  arXiv:hep-th/9602052.
  %%CITATION = HEP-TH/9602052;%%



\bibitem{Dabholkar:1996pc}
  A.~Dabholkar and J.~Park,
  ``Strings on Orientifolds,''
  Nucl.\ Phys.\  B {\bf 477}, 701 (1996)
  [arXiv:hep-th/9604178].
  %%CITATION = NUPHA,B477,701;%%




\bibitem{Grimm:2004uq}
  T.~W.~Grimm and J.~Louis,
  ``The effective action of N = 1 Calabi-Yau orientifolds,''
  Nucl.\ Phys.\  B {\bf 699}, 387 (2004)
  [arXiv:hep-th/0403067].
  %%CITATION = NUPHA,B699,387;%%


%\cite{Strominger:1990pd}
\bibitem{Strominger:1990pd}
  A.~Strominger,
  ``Special Geometry,''
  Commun.\ Math.\ Phys.\  {\bf 133} (1990) 163.
  %%CITATION = CMPHA,133,163;%%

\bibitem{Ferrara:1995ih}
  S.~Ferrara, R.~Kallosh and A.~Strominger,
  ``N=2 extremal black holes,''
  Phys.\ Rev.\  D {\bf 52}, 5412 (1995)
  [arXiv:hep-th/9508072].
  %%CITATION = PHRVA,D52,5412;%%

\bibitem{Ferrara:1997tw}
  S.~Ferrara, G.~W.~Gibbons and R.~Kallosh,
  ``Black holes and critical points in moduli space,''
  Nucl.\ Phys.\  B {\bf 500}, 75 (1997)
  [arXiv:hep-th/9702103].
  %%CITATION = NUPHA,B500,75;%%

%\cite{Moore:1998pn}
\bibitem{Moore:1998pn}
  G.~W.~Moore,
  ``Arithmetic and attractors,''
  arXiv:hep-th/9807087.
  %%CITATION = HEP-TH/9807087;%%

%\cite{Denef:2001xn}
\bibitem{Denef:2001xn}
  F.~Denef, B.~R.~Greene and M.~Raugas,
  ``Split attractor flows and the spectrum of BPS D-branes on the quintic,''
  JHEP {\bf 0105}, 012 (2001)
  [arXiv:hep-th/0101135].
  %%CITATION = JHEPA,0105,012;%%


\bibitem{Shmakova:1996nz}
  M.~Shmakova,
  ``Calabi-Yau Black Holes,''
  Phys.\ Rev.\  D {\bf 56}, 540 (1997)
  [arXiv:hep-th/9612076].
  %%CITATION = PHRVA,D56,540;%%

%\cite{Denef:2000ar}
\bibitem{Denef:2000ar}
  F.~Denef,
  ``On the correspondence between D-branes and stationary supergravity
  solutions of type II Calabi-Yau compactifications,''
  arXiv:hep-th/0010222.
  %%CITATION = HEP-TH 0010222;%%


%\cite{Seiberg:1996nz}
\bibitem{Seiberg:1996nz}
N.~Seiberg and E.~Witten,
``Gauge dynamics and compactification to three dimensions,''
arXiv:hep-th/9607163.
%%CITATION = HEP-TH/9607163;%%

%\cite{Sen:1997pr}
\bibitem{Sen:1997pr}
  A.~Sen,
  ``Strong coupling dynamics of branes from M-theory,''
  JHEP {\bf 9710}, 002 (1997)
  [arXiv:hep-th/9708002].
  %%CITATION = JHEPA,9710,002;%%

%\cite{Witten:1982fp}
\bibitem{Witten:1982fp}
  E.~Witten,
  ``An SU(2) anomaly,''
  Phys.\ Lett.\  B {\bf 117}, 324 (1982).
  %%CITATION = PHLTA,B117,324;%%

%\cite{Uranga:2000xp}
\bibitem{Uranga:2000xp}
  A.~M.~Uranga,
  ``D-brane probes, RR tadpole cancellation and K-theory charge,''
  Nucl.\ Phys.\  B {\bf 598}, 225 (2001)
  [arXiv:hep-th/0011048].
  %%CITATION = NUPHA,B598,225;%%


\bibitem{mooretexas}
 J.~Distler, D.~Freed and G.~W.~Moore, to appear. See also
 \href{http://golem.ph.utexas.edu:2500/jacques/s5/Orientifolds+and+Twisted+KR+Theory}{http://golem.ph.utexas.edu:2500/jacques/s5/Orientifolds+and+Twisted+KR+Theory}

\bibitem{hori}
 K.~Hori  and D.~Gao,
  to appear. See also
 \href{http://online.kitp.ucsb.edu/online/strings05/hori/oh/01.html}{http://online.kitp.ucsb.edu/online/strings05/hori/oh/01.html}


\bibitem{Cecotti:1992qh}
  S.~Cecotti, P.~Fendley, K.~A.~Intriligator and C.~Vafa,
  ``A New supersymmetric index,''
  Nucl.\ Phys.\  B {\bf 386}, 405 (1992)
  [arXiv:hep-th/9204102].
  %%CITATION = NUPHA,B386,405;%%


\bibitem{Kiritsis:1997gu}
  E.~Kiritsis,
  ``Introduction to non-perturbative string theory,''
  arXiv:hep-th/9708130.
  %%CITATION = HEP-TH/9708130;%%


\bibitem{KS}
 M.~Kontsevich and Y.~Soibelman, ``Stability structures, motivic Donaldson-Thomas invariants and cluster transformations,'' arXiv:0811.2435 [math.AG].

\bibitem{Gaiotto:2008cd}
  D.~Gaiotto, G.~W.~Moore and A.~Neitzke,
  ``Four-dimensional wall-crossing via three-dimensional field theory,''
  arXiv:0807.4723 [hep-th].
  %%CITATION = ARXIV:0807.4723;%%

%\cite{Brunner:2004zd}
\bibitem{Brunner:2004zd}
  I.~Brunner, K.~Hori, K.~Hosomichi and J.~Walcher,
  ``Orientifolds of Gepner models,''
  JHEP {\bf 0702}, 001 (2007)
  [arXiv:hep-th/0401137].
  %%CITATION = JHEPA,0702,001;%%


%\cite{LopesCardoso:1998wt}
\bibitem{LopesCardoso:1998wt}
  G.~Lopes Cardoso, B.~de Wit and T.~Mohaupt,
  ``Corrections to macroscopic supersymmetric black-hole entropy,''
  Phys.\ Lett.\ B {\bf 451} (1999) 309
  [arXiv:hep-th/9812082].
  %%CITATION = HEP-TH 9812082;%%

%\cite{LopesCardoso:1999cv}
\bibitem{LopesCardoso:1999cv}
  G.~Lopes Cardoso, B.~de Wit and T.~Mohaupt,
  ``Deviations from the area law for supersymmetric black holes,''
  Fortsch.\ Phys.\  {\bf 48} (2000) 49
  [arXiv:hep-th/9904005].
  %%CITATION = HEP-TH 9904005;%%

%\cite{LopesCardoso:1999xn}
\bibitem{LopesCardoso:1999xn}
  G.~Lopes Cardoso, B.~de Wit and T.~Mohaupt,
  ``Area law corrections from state counting and supergravity,''
  Class.\ Quant.\ Grav.\  {\bf 17} (2000) 1007
  [arXiv:hep-th/9910179].
  %%CITATION = HEP-TH 9910179;%%

%\cite{Braun:2008ua}
\bibitem{Braun:2008ua}
  A.~P.~Braun, A.~Hebecker and H.~Triendl,
   ``D7-Brane Motion from M-Theory Cycles and Obstructions in the Weak Coupling
  Limit,''
  Nucl.\ Phys.\  B {\bf 800}, 298 (2008)
  [arXiv:0801.2163 [hep-th]].
  %%CITATION = NUPHA,B800,298;%%

%\cite{Brunner:2008bi}
\bibitem{Brunner:2008bi}
  I.~Brunner and M.~Herbst,
  ``Orientifolds and D-branes in N=2 gauged linear sigma models,''
  arXiv:0812.2880 [hep-th].
  %%CITATION = ARXIV:0812.2880;%%


%\cite{Candelas:1990rm}
\bibitem{Candelas:1990rm}
  P.~Candelas, X.~C.~De La Ossa, P.~S.~Green and L.~Parkes,
  ``A pair of Calabi-Yau manifolds as an exactly soluble superconformal
  theory,''
  Nucl.\ Phys.\  B {\bf 359}, 21 (1991).
  %%CITATION = NUPHA,B359,21;%%

%\cite{Brunner:1999jq}
\bibitem{Brunner:1999jq}
  I.~Brunner, M.~R.~Douglas, A.~E.~Lawrence and C.~Romelsberger,
  ``D-branes on the quintic,''
  JHEP {\bf 0008}, 015 (2000)
  [arXiv:hep-th/9906200].
  %%CITATION = JHEPA,0008,015;%%

%\cite{Strominger:1995cz}
\bibitem{Strominger:1995cz}
  A.~Strominger,
  ``Massless black holes and conifolds in string theory,''
  Nucl.\ Phys.\  B {\bf 451}, 96 (1995)
  [arXiv:hep-th/9504090].
  %%CITATION = NUPHA,B451,96;%%

%\cite{Collinucci:2008ht}
\bibitem{Collinucci:2008ht}
  A.~Collinucci and T.~Wyder,
  ``The elliptic genus from split flows and Donaldson-Thomas invariants,''
  arXiv:0810.4301 [hep-th].
  %%CITATION = ARXIV:0810.4301;%%


 %\cite{Denef:2007yt}
\bibitem{Denef:2007yt}
  F.~Denef, D.~Gaiotto, A.~Strominger, D.~Van den Bleeken and X.~Yin,
  ``Black hole deconstruction,''
  arXiv:hep-th/0703252.
  %%CITATION = HEP-TH/0703252;%%


%\cite{Bena:2007qc}
\bibitem{deconstr}

   S.~D.~Mathur,
  ``The fuzzball proposal for black holes: An elementary review,''
  Fortsch.\ Phys.\  {\bf 53}, 793 (2005)
  [arXiv:hep-th/0502050].
  %%CITATION = FPYKA,53,793;%%

 I.~Bena, C.~W.~Wang and N.~P.~Warner,
  ``Plumbing the Abyss: Black Ring Microstates,''
  JHEP {\bf 0807}, 019 (2008)
  [arXiv:0706.3786 [hep-th]].
  %%CITATION = JHEPA,0807,019;%%

  V.~Balasubramanian, E.~G.~Gimon and T.~S.~Levi,
  ``Four Dimensional Black Hole Microstates: From D-branes to Spacetime Foam,''
  JHEP {\bf 0801}, 056 (2008)
  [arXiv:hep-th/0606118].
  %%CITATION = JHEPA,0801,056;%%

  J.~de Boer, S.~El-Showk, I.~Messamah and D.~V.~d.~Bleeken,
  ``Quantizing N=2 Multicenter Solutions,''
  arXiv:0807.4556 [hep-th].
  %%CITATION = ARXIV:0807.4556;%%

%\cite{Gaiotto:2006ns}
\bibitem{Gaiotto:2006ns}
  D.~Gaiotto, A.~Strominger and X.~Yin,
  ``From AdS(3)/CFT(2) to black holes / topological strings,''
  JHEP {\bf 0709}, 050 (2007)
  [arXiv:hep-th/0602046].
  %%CITATION = JHEPA,0709,050;%%

%\cite{deBoer:2006vg}
\bibitem{deBoer:2006vg}
  J.~de Boer, M.~C.~N.~Cheng, R.~Dijkgraaf, J.~Manschot and E.~Verlinde,
  ``A farey tail for attractor black holes,''
  JHEP {\bf 0611}, 024 (2006)
  [arXiv:hep-th/0608059]. 
  
%\cite{Gaiotto:2006wm}
\bibitem{Gaiotto:2006wm}
  D.~Gaiotto, A.~Strominger and X.~Yin,
  ``The M5-brane elliptic genus: Modularity and BPS states,''
  JHEP {\bf 0708}, 070 (2007)
  [arXiv:hep-th/0607010].
  %%CITATION = JHEPA,0708,070;%%



\bibitem{Nis4}

 %\cite{Sen:2007vb}
  A.~Sen,
  ``Walls of Marginal Stability and Dyon Spectrum in N=4 Supersymmetric
  String Theories,''
  JHEP {\bf 0705}, 039 (2007)
  [arXiv:hep-th/0702141].
  %%CITATION = JHEPA,0705,039;%%

 %\cite{Cheng:2007ch}
  M.~C.~N.~Cheng and E.~Verlinde,
  ``Dying Dyons Don't Count,''
  JHEP {\bf 0709}, 070 (2007)
  [arXiv:0706.2363 [hep-th]].
  %%CITATION = JHEPA,0709,070;%%


%\cite{Sen:2008ta}
  A.~Sen,
  ``N=8 Dyon Partition Function and Walls of Marginal Stability,''
  JHEP {\bf 0807}, 118 (2008)
  [arXiv:0803.1014 [hep-th]].
  %%CITATION = JHEPA,0807,118;%%


%\cite{Sen:2008ht}
  A.~Sen,
  ``Wall Crossing Formula for N=4 Dyons: A Macroscopic Derivation,''
  JHEP {\bf 0807}, 078 (2008)
  [arXiv:0803.3857 [hep-th]].
  %%CITATION = JHEPA,0807,078;%%


  M.~C.~N.~Cheng and E.~P.~Verlinde,
  ``Wall Crossing, Discrete Attractor Flow, and Borcherds Algebra,''
  arXiv:0806.2337 [hep-th].
  %%CITATION = ARXIV:0806.2337;%%

  
   %\cite{Cheng:2008kt}
  M.~C.~N.~Cheng and A.~Dabholkar,
  ``Borcherds-Kac-Moody Symmetry of N=4 Dyons,''
  arXiv:0809.4258 [hep-th].
  %%CITATION = ARXIV:0809.4258;%%

  M.~C.~N.~Cheng and L.~Hollands,
  ``A Geometric Derivation of the Dyon Wall-Crossing Group,''
  arXiv:0901.1758 [hep-th].
  %%CITATION = ARXIV:0901.1758;%%



%\cite{Dijkgraaf:1996it}
\bibitem{Dijkgraaf:1996it}
  R.~Dijkgraaf, E.~P.~Verlinde and H.~L.~Verlinde,
  ``Counting Dyons in N=4 String Theory,''
  Nucl.\ Phys.\  B {\bf 484}, 543 (1997)
  [arXiv:hep-th/9607026].
  %%CITATION = NUPHA,B484,543;%%

%\cite{Aganagic:2005dh}
\bibitem{Aganagic:2005dh}
  M.~Aganagic, A.~Neitzke and C.~Vafa,
  ``BPS microstates and the open topological string wave function,''
  arXiv:hep-th/0504054.
  %%CITATION = HEP-TH/0504054;%%







\end{thebibliography}
\end{document}